%% file: delta_paper.tex
\documentclass[sigconf]{acmart}

\usepackage{booktabs} 

\usepackage{balance}

\usepackage{algorithm}
\usepackage{algpseudocode}

\usepackage{array}

\usepackage{colortbl}

\usepackage{enumitem}

\newcommand{\vect}[1]{\mathbf{#1}}
\newcommand{\tuple}[1]{\langle {#1} \rangle}
\DeclareMathOperator{\Real}{\mathbb{R}}

\newcolumntype{$}{>{\global\let\currentrowstyle\relax}}
\newcolumntype{^}{>{\currentrowstyle}}
\newcommand{\rowstyle}[1]{\gdef\currentrowstyle{#1}%
	#1\ignorespaces
}

\newif\ifforconf

\forconffalse

\begin{document}

%
\ifforconf
%
%
\copyrightyear{2018}
\acmYear{2018} 
\setcopyright{iw3c2w3}
\acmConference[WWW 2018]{The 2018 Web Conference}{April 23--27, 2018}{Lyon, France}
\acmBooktitle{WWW 2018: The 2018 Web Conference, April 23--27, 2018, Lyon, France}
\acmPrice{}
\acmDOI{10.1145/3178876.3186049}
\acmISBN{978-1-4503-5639-8/18/04}
\fancyhead{}
\else
%
%
\copyrightyear{2018}
\acmYear{2018} 
\setcopyright{iw3c2w3}
\acmConference[To appear in WWW 2018]{To appear in WWW}{2018}{Lyon, France}
\acmBooktitle{To appear in: WWW 2018, The 2018 Web Conference, April 23--27, 2018, Lyon, France}
\acmPrice{}
\fancyhead[LE,RO]{\em To appear in: WWW 2018, The 2018 Web Conference}
\fi

\title{A Fast Deep Learning Model for Textual Relevance \\ in Biomedical Information Retrieval}

\author{Sunil Mohan}
\authornote{Corresponding author. This work was done while this author was at NCBI.}
\affiliation{%
	\institution{Chan Zuckerberg Initiative}
	\city{Palo Alto} 
	\state{CA} 
	\postcode{94301}
	\country{USA}
}
\email{smohan@chanzuckerberg.com}

\author{Nicolas Fiorini}
\affiliation{%
	\institution{National Center for Biotechnology Information}
	\city{Bethesda} 
	\state{MD} 
	\postcode{20894}
	\country{USA}
}
\email{nicolas.fiorini@nih.gov}

\author{Sun Kim}
\affiliation{%
	\institution{National Center for Biotechnology Information}
	\city{Bethesda} 
	\state{MD} 
	\postcode{20894}
	\country{USA}
}
\email{sun.kim@nih.gov}

\author{Zhiyong Lu}
\affiliation{%
	\institution{National Center for Biotechnology Information}
	\city{Bethesda} 
	\state{MD} 
	\postcode{20894}
	\country{USA}
}
\email{zhiyong.lu@nih.gov}

\renewcommand{\shortauthors}{S. Mohan et al.}

\begin{abstract}
Publications in the life sciences are characterized by a large technical vocabulary, with many lexical and semantic variations for expressing the same concept. Towards addressing the problem of relevance in biomedical literature search, we introduce a deep learning model for the relevance of a document's text to a keyword style query. Limited by a relatively small amount of training data, the model uses pre-trained word embeddings. With these, the model first computes a variable-length Delta matrix between the query and document, representing a difference between the two texts, which is then passed through a deep convolution stage followed by a deep feed-forward network to compute a relevance score. This results in a fast model suitable for use in an online search engine. The model is robust and outperforms comparable state-of-the-art deep learning approaches.
\end{abstract}

%
%



\maketitle

\input{delta_body}

\bibliographystyle{ACM-Reference-Format}
\balance
\bibliography{delta_bib} 

\end{document}

%% file: delta_body.tex
\section{Introduction}

PubMed\textsuperscript{\textregistered}\footnote{\url{http://pubmed.gov}} is a free online search engine covering over 27 million articles from biomedical and life sciences journals and other texts, with about 1 million added each year. It is used worldwide by biomedical researchers, add healthcare professionals as well as lay people, serving about 3 million queries a day \cite{Dogan:2009}. While expert users either search for most recent articles by an author or construct elaborate query expressions, most queries are short keyword-ese, covering one or two biomedical concepts. Although the size of the corpus is much smaller than in general web search, biomedical literature uses a very large technical vocabulary (e.g. the UMLS\footnote{\url{http://umlsks.nlm.nih.gov}} metathesaurus \cite{Bodenreider:2004} specifies over 3 million biomedical concepts, along with several lexical variations and synonymous phrases. This makes it much harder to identify concepts across documents (e.g. see \cite{Kim-etal:2015}). To improve the retrieval, PubMed expands a user's query by mapping it to related MeSH\textsuperscript{\textregistered} terms \cite{Lu:2009}. While this increases recall, it often decreases precision \cite{Hersh-etal:2000}. Usage analysis \cite{Dogan:2009} shows that PubMed users are persistent, often reformulating their query, issuing over 4 queries per session on average. As part of improving relevance for such keyword queries, we describe a deep learning model that addresses the relevance of a document's text to the query. The eventual goal is for this model to be incorporated as a factor into a reranker that also includes other document attributes and metadata (e.g. year, journal).

To train our model, we collected data from PubMed click logs, restricting this to relevance search instead of the default sort order by date. Removing author searches and disjunctive boolean expressions resulted in a training set of about 20k queries. Given the small size of this data, we pre-trained word embeddings using \texttt{word2vec} \cite{Mikolov:2013:NIPS} on the entire PubMed corpus, producing a vocabulary of about 200k. This large gap beween training data and vocabulary sizes highlights a major challenge: how to make the model robust? Our Delta deep learning model begins by computing a variable-sized `Delta' matrix between a document and a query, comprising the vector difference between document word embeddings and the closest matching query word, and three scalar similarity measures between the query and document. The document is truncated to control run-time cost. The Delta matrix is processed through a stacked convolutional network and pooled to a fixed length. This, together with a summary query match statistic, is processed by a feed-forward network to produce a relevance score. Pairwise loss is optimized for training. This approach produces a model that is both robust, and fast enough for use in a search engine.

In addition to model robustness, we also wanted to address two common search engine problems: (i) the {\em under-specified query problem} \cite{Durao-etal:2013}, where even irrelevant documents have prominent presence of the query terms, and relevance requires analysis of the topics and semantics not directly specified in the query, and (ii) the {\em term mismatch problem} \cite{Furnas-et-al:1987}, which requires detection of related alternative terms or phrases in the document when the actual query terms are not in the document. Our experiments show the Delta model outperforms traditional lexical match factors and some related state-of-the-art neural approaches.

The next sections discuss some related work, followed by a description of the model, the experiments and evaluation of results, ending with some concluding remarks.

\section{Related Work}

Traditional lexical Information Retrieval (IR) factors, like Okapi BM25 \cite{Robertson:1994} and Query Likelihood \cite{Miller:1999}, measure the prominence of query terms occurring in documents treated as bags of words. Neural approaches to text relevance attempt to go beyond exact matches of query terms in documents, and model a degree of semantic match as a complex function in a continuous space (good reviews can be found in \cite{Zhang-etal:2016:NeuralIR-Review, Mitra-Craswell:2017:NeuralIR}). We will discuss some related approaches here.

Most neural models begin by mapping words to points embedded in a real space. A popular approach (e.g. \cite{Hu-et-al:2014:NIPS,Guo-et-al:2016}), also used in our model, is to pre-train word embeddings, e.g. using \texttt{word2vec} \cite{Mikolov:2013:ICLR, Mikolov:2013:NIPS}. The benefit of this approach is that a much larger unlabeled corpus can be used to train the embeddings, and our `Delta matrix' takes advantage of the semantic relationships captured in the vector differences between words.

The simplest embeddings based model is Word Mover's Distance (WMD) \cite{Kusner:2015:ICML}, a non-parameterized model for text similarity that does not require any training. We use this as one of our baselines.  Pre-trained embeddings are not necessarily targeted for optimal relevance scores. Nalisnick et al. \shortcite{Nalisnick-etal:2016} also use the `input' vectors normally discarded by \texttt{word2vec} to overcome some of this limitation. The `DSSM' models of \cite{Huang:2013,Shen:2014} take a different approach by mapping each word to a bag of letter tri-grams and combining the corresponding one-hot vectors. Xiong et al. \shortcite{Xiong-etal:2017} show that training word embeddings as part of the relevance model has a major impact on the performance of a relvance model. However this requires a large amount of training data. Diaz et al. \shortcite{Diaz-etal:2016} show that `locally trained' embeddings on pseudo-relevance feedback documents can provide better results, while admitting that this approach is not ``computationally convenient''.

Neural relevance models also differ in how they process document and query text. Some (e.g. \cite{Huang:2013,Shen:2014,Gao-et-al:2014:EMNLP,Hu-et-al:2014:NIPS}) process each document and query using separate `Siamese' networks into independent semantic vectors. A second stage then scores the similarity between these vectors. This approach is very attractive for search engines, because the document vectors can be pre-processed and stored, and the query vector need be produced once before scoring the documents, significantly reducing the cost at query time. We use the recent model described in \cite{Severyn:2015} as a baseline. 

Another approach to text matching first develops `local interactions' by comparing all possible combinations of words and word sequences between the document and query texts, often starting with a document-query word similarity matrix. Examples are described in \cite{Hu-et-al:2014:NIPS, Lu:Li:2013:NIPS, Guo-et-al:2016, Pang-etal:2016, Xiong-etal:2017}. The authors in \cite{Guo-et-al:2016} argue that the local interaction based approach is better at capturing detail, especially exact query term matches, and in their experiments their `DRMM' model outperforms many previous approaches. This is a more computationally intensive architecture that does not allow any pre-computation. We take a similar approach by pairing each document word with a single query word, followed by deep convolutions to capture some related compositional semantics. Run-time cost in our approach is controlled by truncating the document. We show that our approach outperforms the DRMM model.

The `PACRR-firstk' model in \cite{Hui-etal:2017:EMNLP} also truncates the document, then processes the resulting similarity matrix through several 2D convolutional layers aimed at capturing n-gram similarities, followed by a recurrent layer, resulting in a fairly complex model. The `DUET' model described in \cite{Mitra-etal:2017:WWW} combines a local interaction model with an independent semantic vector model, with the goal of combining the benefits of `exact match' and embedding based word similairities. Our simpler approach explicitly targets run-time efficiency, and a variant of the Delta model combines some lexical factors (similar to \cite{Severyn:2015}) to further improve ranking performance.

\section{The Delta Model}

\begin{figure}
	\includegraphics[width=\columnwidth]{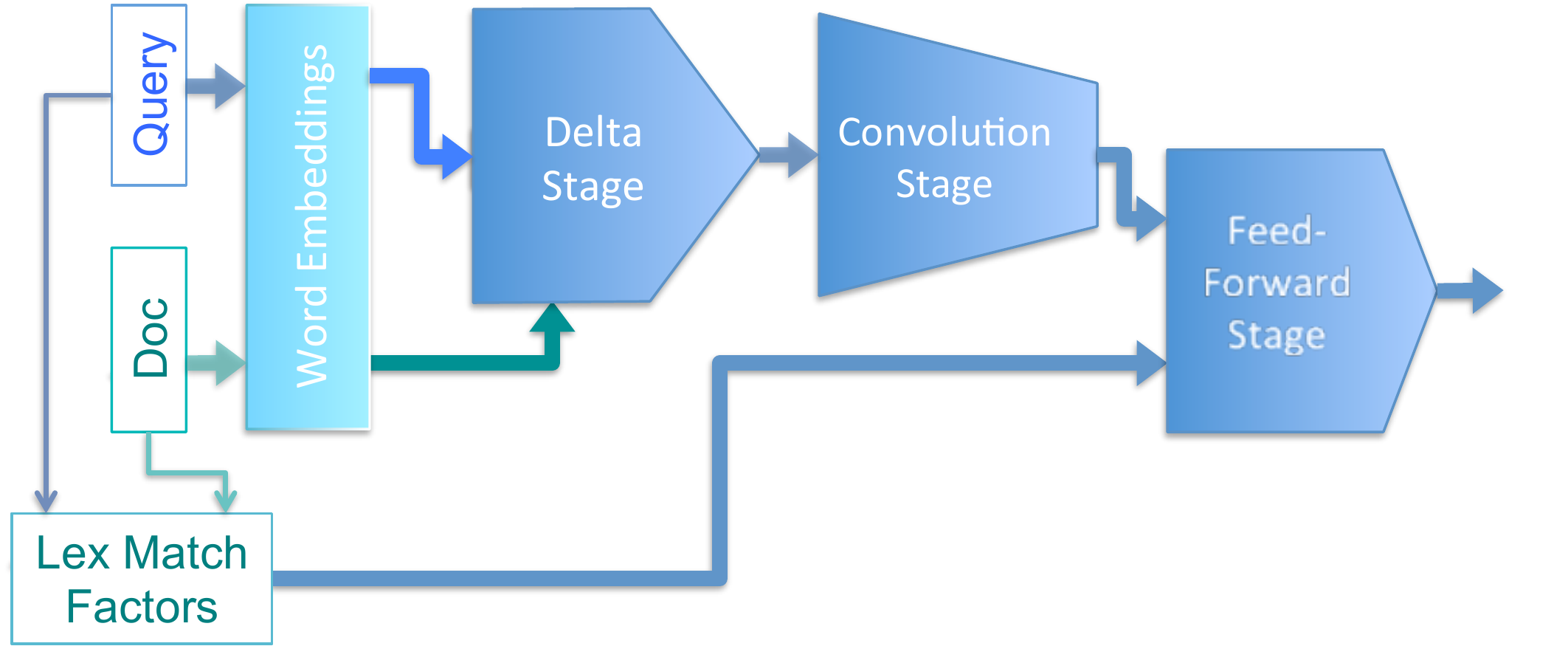}
	\caption{The Delta Relevance Model.}
	\label{fig:DeltaArch}
\end{figure}

The components of the Delta Relevance Model (figure~\ref{fig:DeltaArch}) are described below. The unshaded blocks represent inputs to the model: two vectors of word indices, one each for the Document $D$ and the Query $Q$, and a vector of query-document Lexical Match factors $\vect{L}_{DQ} = \text{lexmatch}(D, Q)$ for some chosen lexical match function.

The small size of the training data ($\sim 20,000$ queries) compared to the vocabulary size ($\sim 200,000$) prevented us from training word embeddings as part of the model training. We had to adapt the word vectors pre-trained using word2vec's unsupervised approach, to the task of relevance prediction. The Delta model uses two techniques that help in this and thus learn a richer and more robust decision surface. Changing the input space from word embeddings to differences in word embeddings shifts the domain of the decision surface to coordinates relative to the query. In addition, the Delta model's use of a stack of convolution layers instead of a single layer adds more non-linearities to help capture a complex decision surface, a technique successful in image recognition \cite{Urban-etal:2017}. The convolution layers also extract relevance-match signals from text $n$-grams, and are much faster than a recurrent layer which has a similar goal.

\subsection{Word Embeddings}

We leveraged the large PubMed corpus of over 27 million documents to pre-train the word vectors, using the SkipGram Hierarchical Softmax method of \texttt{word2vec} \cite{Mikolov:2013:NIPS}, with a window size of $\pm 5$, a minimum term-frequency of 101, and a word-vector size of $V = 300$ (see \cite{Chiu-etal:2016:BioNLP} for experiments with different parameter settings for biomedical text). This resulted in a vocabulary of 207,716 words. Rare words were replaced with the generic ``UNK'' token, which was initialized to $\sim U[-0.25, 0.25]$, as in \cite{Severyn:2015}.

Given a document word sequence $D = \tuple{w^{d}_1, \ldots, w^{d}_N}$ and query text $Q = \tuple{w^{q}_1, \ldots, w^{q}_M}$, where $w_i$ are indices into the vocabulary, the Embeddings layer replaces each word with its vector, giving us $D^{e} = \tuple{d_1, \ldots, d_N}, Q^{e} = \tuple{q_1, \ldots, q_M}$ where each $d_i, q_i \in \Real^{V}$, and $V$ is the size of the word embedding. If a document has fewer than $N$ words, or the query fewer than $M$ words, they are padded on the right with zeros. Longer documents are truncated, and $M$ is the longest query length in our data (see section \ref{sec:TheData}).

\subsection{The Delta Stage}

This is an unparameterized stage, responsible for computing the Delta Matrix between the Document and the Query as follows:
\begin{enumerate}[leftmargin=2em]
	\item Compute the Euclidean distance between each pair $d_i, q_j$.
	
	\item For each document word $d_i$, determine the closest query word $q^{*}_i$, using these distances.
	
	\item Compute the vector differences $(d_i - q^{*}_i)$.
	
	\item Compute the Delta features: cosine$(d_i, q^{*}_i)$, $|d_i - q^{*}_i|$ and normalized proximity similarity metric $1 - |d_i - q^{*}_i| / (|d_i| + |q^{*}_i|)$.
\end{enumerate}
The output of this stage is $\Delta = \tuple{\delta_1, \ldots, \delta_N}$, a $N \times (V + 3)$ sized real matrix. All operations above are masked to ignore padding.

\subsection{The Network}

The trainable portion of the Delta model consists of a Convolutional Stage followed by a Feed-Forward Stage. The Convolutional stage attempts to pick up significant contextual and $n$-gram similarity features. These are combined with the Lexical Match features $\vect{L}_{DQ}$, then processed by the Feed-Forward Stage to produce a final relevance score. The weights for the layers in these stages comprise the trainable parameters for the model.

A convolution operation \cite{LeCun:1989,Goodfellow-et-al:2016} has the parameters: border mode, number of filters or feature maps $n_f$, filter width $k$ and stride $s$. We use 1-dimensional convolution along the text width with border mode `{\em same}' which implicitly pads the input on either side before convolving, and a stride $s=1$, resulting in an output of the same width as the input. Using $conv(X; n_f, k)$ to represent such a convolution on input $X$, the Delta model's Convolutional Stage performs the following operations:
\begin{align*}
\vect{Y}^{(i)} &= f_i(conv(\vect{Y}^{(i-1)}; n_f, k)) & i = 1, \ldots, N_C
\end{align*}
where $f_i$ is an activation function, $\vect{Y}^{(0)} = \Delta$, $N_C$ is the number of convolution layers, and $\vect{Y}^{(N_C)} \in \Real^{N \times n_f}$. The number of filters and filter width are kept the same for each layer, and all operations are masked to ignore padding.

The output of the final convolution layer is `max-pooled' by taking the maximum of each of the $n_f$ output features along the text width dimension, yielding a vector of size $n_f$. This is the output of the Convolution Stage:
\[
\vect{Y} = \max(\vect{Y}^{(N_C)}; \text{axis}=0) \in \Real^{n_f}
\]
This output is combined with the Lexical Match features $\vect{L}_{DQ}$ and sent to the Feed-Forward stage which is a series of $N_F$ layers:
\begin{align*}
\vect{Z}^{(i)} &= g_i(\vect{Z}^{(i-1)} \cdot \vect{W}_i + \vect{b}_i) & i = 1, \ldots, N_F
\end{align*}
Here: $g_i$ is an activation function, $\vect{Z}^{(0)} = \tuple{\vect{Y}, \vect{L}_{DQ}}$,\hspace{1ex} $\vect{Z}^{(i-1)} \cdot \vect{W}_i$ is the matrix multiplication operation, $N_F$ is the number of layers in the Feed-Forward stage, and $\vect{W}_i, \vect{b}_i$ are sized so that the number of outputs of the $i$-th layer are $K_i$. The output of the final layer is a single number, i.e. $K_{N_F} = 1, \vect{Z}^{(N_F)} \in \Real$, representing the relevance score of the document $D$ to the query $Q$.

\subsection{Training The Model}

The training data derived from PubMed's click logs provides relevance levels for query-document pairs based on the number of clicks they received (see next section for more details). The Delta relevance model is trained to give more relevant documents a higher score by tuning its parameters $\Theta$ to minimize the pairwise maximum margin loss. Given a query $Q$ and two matching documents $D^+, D^-$ where $D^+$ has higher relevance to the query than $D^-$, the loss for this triple is expressed as:
\[
\mathcal{L}(Q, D^+, D^-; \Theta) = \max(0, 1 - s(Q, D^+; \Theta) + s(Q, D^-; \Theta))
\]
where $s(Q, D; \Theta)$ is the relevance score produced by the Delta model for the query-document pair $Q, D$. The Adagrad \cite{Duchi:2011} stochastic gradient descent method was used to train the model, using a mini-batch size of 256. In addition to early stopping, separate L2 regularization costs were added on the weights of the Convolutional and Feed-Forward stages, and a dropout layer was added before the max-pooling layer in the Convolutional stage. The regularization coefficients and dropout probability were tuned using the held out validation data. Adding dropout layers to the Feed-Forward stage was also tested but was not found to help.

\section{Experimental Setup}

\begin{table*}
	\small
	\begin{center}
		\begin{tabular}{lrrrr}
			\toprule
			{} & Full Test Data & Neg20+ & OneNewWord & AllNewWords\\
			\midrule
			Nbr. of Queries &   6,734 &   2,600 &   1,732 &     933 \\
			Nbr. of Samples & 413,971 & 208,723 &  86,438 &  47,825 \\
			Prop. of Samples +ive & 45.2\% & 39.5\% & 49.2\% & 49.2\% \\
			Prop. of Samples -ive & 54.8\% & 60.5\% & 50.8\% & 50.8\% \\
			+ives without all Query terms in Title & 38.7\% & 13.9\% & 32.7\% & 22.9\%    \\
			-ives with all Query terms in Title    & 59.5\% & 83.6\% & 68.2\% & 78.0\% \\
			\bottomrule
		\end{tabular}
	\end{center}
	\caption{\label{tab:TestData}Test data and its subsets}
\end{table*}

\subsection{The Data}
\label{sec:TheData}

We collected query-document pairs extracted from PubMed click logs over several months where users selected `Best Match' (relevance) as the retrieval sort order and clicked on at least one document in the search results. We recorded the first page of results of up to 20 documents, supplemented with the clicked document if it was not on the first page. Since our primary goal was to improve relevance for simple keyword style queries, we discarded queries containing disjunctive expressions, faceted queries, and queries longer than 7 words. Log extracts were further restricted to queries with at least 21 documents, and at least 3 clicked documents. These filters reduced the logs to about 33,500 queries, which were randomly split to 60\% training, and 20\% each for validation and testing.

{\em Relevance Levels.\ }
The relevance level assigned to each query-document pair extracted from click logs is a probability of relevance, scaled to the range [0, 100] so that the minimum possible non-zero relevance is 1. 
For each query-document pair $Q,D$, we accumulated over the collection period the number of click-throughs $c(D,Q)$ from search results to the document summary page, whether the document's full-text was available in PubMed $I_{ft}(D) \in \{0, 1\}$, and the number of subsequent click-throughs $c_{ft}(D,Q)$ to the document's full-text. These were used to derive a weighted click-count $c_w(Q,D)$ that rewarded documents for which full-text was requested without penalizing those for which full-text was not available. From that a probability of relevance $rel(Q,D)$ was calculated:
\begin{align*}
c_w(Q,D) &= \left(\mu + (1 - I_{ft}(D)).\lambda\right) \cdot c(D,Q) + (1 - \mu) \cdot c_{ft}(D,Q)  \\
rel(Q,D) &= \frac{c_w(Q,D)}{\sum_{D^{'}} c_w(Q,D^{'})}  \\
srel(Q,D) &= 1 + 99 \times rel(Q,D) \ldots\text{ if } rel(Q,D) > 0 \\
		  &= 0 \hspace{6.8em} \ldots\text{ if } rel(Q,D) = 0
\end{align*}
Finally, we scaled the non-zero relevance levels to the range $(1, 100]$ to get $srel(Q,D)$. This ensured a minimum margin between documents of low relevance and no relevance, and also put a high penalty in the NDCG metric for ranking high relevance documents below low relevance ones. The coefficients were tuned to match NCBI domain experts' relevance judgments: $\mu = 0.333, \lambda = 0.067$.

{\em Tokenization.\ }
Each document in our data had a Title and an Abstract. For the neural models, we concatenated these to form the document's `Text'. All document and query text was tokenized by splitting on space and punctuation, while preserving abbreviations and numeric forms, followed by a conversion to lower-case. To further reduce the vocabulary size, all punctuation was discarded and numeric forms were collapsed into 7 classes: Integer, Fraction in (0, 1), Real number, year ``19xx'', year ``20xx'', Percentage (number followed by ``\%''), and dollar amount (number preceded by ``\$''). While \texttt{word2vec} processed the tokenzed documents in sentences, the document input to the neural models was a flat sequence of words without sentence breaks or markers. The distribution of document text widths (nbr. of words) in the data is shown in figure~\ref{fig:DocWidthsAndDrmmPerf}a. We experimented with stopword removal in the Query and the Document, but they did not help.

{\em Test Data Subsets.\ }
The 20\% held out test data comprised 6,734 queries and 413,971 samples (query-document pairs). Presence of query words in a document's Title is often a good indication of relevance. Among the relevant documents (``+ives'') for all the test queries, 38.7\% did not contain all query terms in the title. Similarly 59.5\% of all the non-relevant documents (``-ives'') actually contained all the query terms in their title (see table~\ref{tab:TestData}).

In addition to comparing ranking metrics of the different approaches on the test data, we wanted to explore model robustness, and model performance with {\em under-specified queries}. To help answer these questions, we also compared ranking metrics on the following subsets of the test data:
\begin{description}
	\item[{\textbf Neg20+}:] This consisted of all queries for which there were at least 20 non-relevant documents that contained all the query words in the title. This subset was used to evaluate performance on under-specified queries.
	
	\item[{\textbf OneNewWord}:] The 1,732 test queries that contained at least one new word not occurring in any training or validation queries.
	
	\item[{\textbf AllNewWords}:] A smaller subset of queries all of whose words were new: none of the training or validation queries included these words.
\end{description}

The last two subsets help evaluate model robustness. The statistics of the test data and its subsets are summarized in table~\ref{tab:TestData}.

\subsection{Configuration Settings for the Delta Model}

The Delta Model's hyper-parameters were tuned to optimize the ranking metric NDCG.20 on the validation data. We found truncating documents to the first $N = 50$ words provided a good compromise between ranking performance and the run time to score a query-document pair (discussed below), with larger values providing only marginal improvements. The maximum query size was $M = 7$ as described above. The Convolutional stage used $N_C = 3$ layers of convolutions, each with  a filter width $k = 3$. We report metrics for various number of filters $n_f$ below. The Feed-Forward stage used $N_F = 3$ layers. Finally, we found downsampling the training data so that there were an equal number of relevant and non-relevant documents for each query to produce the best model, resulting in 7,084,244 training samples of $(Q, D^+, D^-)$ triples. This downsampling was also performed for the other neural models described below. The validation and test data were not downsampled.

With the maximum-margin loss function, there was no reason to constrain the range of the final layer's activation function. We got best results using the Leaky Rectified Linear Unit (Leaky ReLU)  \cite{Maas-etal:2013} with the slope of the negative region fixed at $\alpha = 0.3$. The Leaky ReLU was also used as the activation function for all the other layers of the Feed-Forward and Convolutional stages.

An earlier version of the Delta model is described in \cite{Mohan-etal:2017:BioNLP17}. The main changes since then are: a simpler Delta Matrix, changes to the activation functions used in all the stages, training to a pairwise loss function with different sample weighting, and comprehensive test on a number of lexical features. These changes resulted in a $\sim 10\%$ improvement in the NDCG metrics. We only report metrics for the current version of the Delta model below, along with a comparison against some new baselines.

\subsubsection{Relevance-based Sample Weighting.}

Best results were obtained by adding a weight to each $(Q, D^+, D^-)$ training sample in the loss function by taking the square-root of the difference in the scaled-relevance levels of the two documents:
\[
\text{weight}(Q, D^+, D^-) = (srel(Q, D^+) - srel(Q, D^-))^{0.5}
\]

\subsubsection{Lexical Match Features.}
\label{sec:LexMatchFeatures}

As an extension to the ``word overlap measures'' used in the SevMos model \cite{Severyn:2015}, we tested 18 features for use as the `Lexical Match Factors' input to the Delta model:
\begin{enumerate}[leftmargin=2.2em]
	\item Proportion of unique Query words present in document Text.
	\item Proportion of unique Query bigrams present in doc Text.
	\item Jaccard Similarity between Query and document Text.
	\item IDF weighted version of (1).
	\item An IDF weighted version of Jaccard Similarity (3).
	
	\item BM25 on Query, document Title.
	\item BM25 on Query, document Abstract.
	\item BM25 on Query, document Text.
	
	\item Proportion of unique Query words present in document Title.
	\item Proportion of unique Query bigrams present in doc Title.
	\item Jaccard Similarity between Query and document Title.
	\item IDF weighted version of (9).
	\item An IDF weighted version of Jaccard Similarity (11).
	
	\item Proportion of unique Query words present in doc Abstract.
	\item Proportion of unique Query bigrams present in doc Abstract.
	\item Jaccard Similarity between Query and document Abstract.
	\item IDF weighted version of (14).
	\item An IDF weighted version of Jaccard Similarity (16).
\end{enumerate}

To compute these factors, Queries and Documents were tokenized as described above, without the rare word conflation needed for computing word embeddings. Document Text refers to the combined Title and Abstract, each of these (as well as the Query) treated as a sequence of words with no truncation. These factors were selected based on the speed of their computation in a search engine. Factors (3, 5, 11, 13, 16, 18) were also used in \cite{Severyn:2015}.

\subsection{Baselines}

We compared the performance of the Delta deep learning model against some traditional bag-of-words based textual relevance factors, a distance measurement based on distributional representations of words, and a couple of recent neural network models.

\subsubsection{Lexical Factors.}

We compared the performance of Okapi BM25 \cite{Robertson:1994} on the document Title, Abstract and Text (Title + Abstract), and found BM25 on Title to give the best ranking performance, with parameter settings at $k_1 = 2.0$ and $b = 0.75$.

The second lexical factor we tested was Unigram Query Likelihood (UQLM), which estimates the probability with which the most likely random process that generated the bag-of-words representation of the document, would generate the query. It is based on a generative unigram language model that is a mixture of two multinomial models \cite{Miller:1999} based on the document and the corpus, combined using Dirichlet smoothing \cite{Lafferty:Zhai:2001, Zhai:Lafferty:2004}. Just like in the case of BM25, we found UQLM applied to the document Title to perform the best, and quote only those metrics below.

\subsubsection{Word Mover's Distance.}

Since all the neural models in our experiments started with pre-trained word embeddings, the {\em Word Mover's Distance} (WMD) model \cite{Kusner:2015:ICML} for text dis-similarity (score decreases with increasing similarity) was an obvious baseline approach. Based on the Earth Mover's Distance \cite{Rubner:1998} applied to a bag-of-words representation for text, it is a non-parameterized approach to determine the minimum amount of total transportation cost (sum of product of inter-word cost and amount transported) needed to convert one document into the other. It uses the Euclidean distance between the word-vector representations of two words as the cost of moving from one word to another. We only report metrics for WMD applied to the document Title without removal of stop-words, as it performed better than the other alternatives tested.

\subsubsection{The Severyn-Moschitti Model.}

As a recent example of the Independent Semantic Vector approach, we implemented the relevance classification model described in \cite{Severyn:2015}, along with a few variations. The query and document are fed into separate  Convolutional stages, each comprising a single convolution layer with 256 feature maps and a filter width of 5, followed by Dropout and Global Max-Pooling. A similarity measure is computed from these pooled outputs using a similarity weight matrix. The similarity measure, the pooled outpus, and some lexical match features (``overlap measures'' in \cite{Severyn:2015}) are fed into a Classifier stage consisting of a series of feed-forward layers. In our experiments, we provided the SevMos models with all 18 lexical match features described in section~\ref{sec:LexMatchFeatures}. Optimal values for the L2-regularization and Dropout probability hyper-parameters were determined by tuning on validation data, as described for the Delta model.

We tested several variants of this model covering: replacing the single convolution layer with a 3-layer stack of convolutions of filter width 3, similar to the Delta model's Convolutional stage; training the model as a classifier v/s a relevance scorer to the pairwise max-margin loss; and various sample-weighting schemes. Best results were obtained with the classification model using a 3-layer convolution stack and square-root weighting of samples. We report the metrics for this approach as the ``SevMos-C3'' model below, and the corresponding single convolution layer based classifier as the ``SevMos-C1'' model.


\begin{figure}
	\includegraphics[width=0.8\columnwidth]{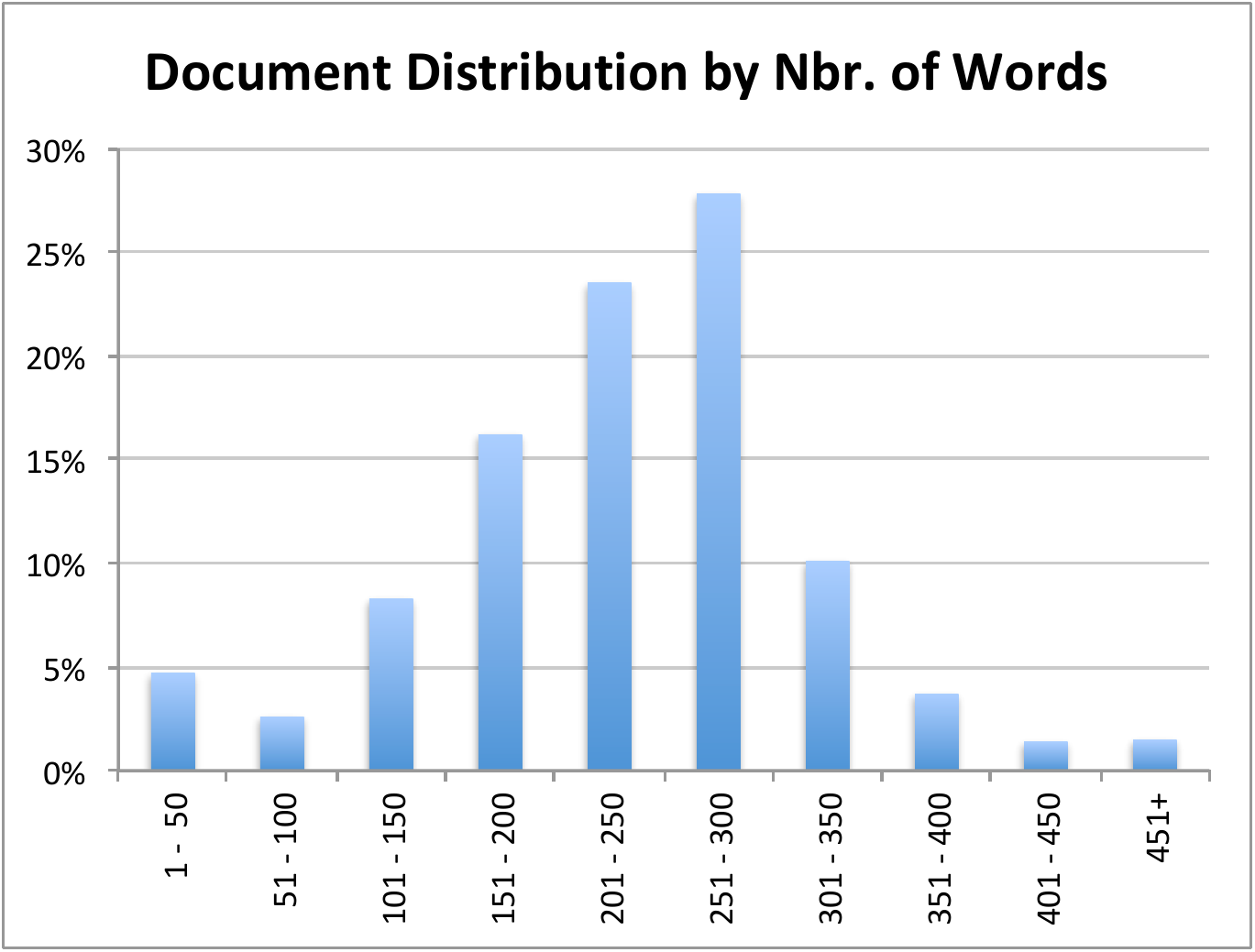}
	{\hspace{4.5em}(a)}\\*
	\includegraphics[width=0.8\columnwidth]{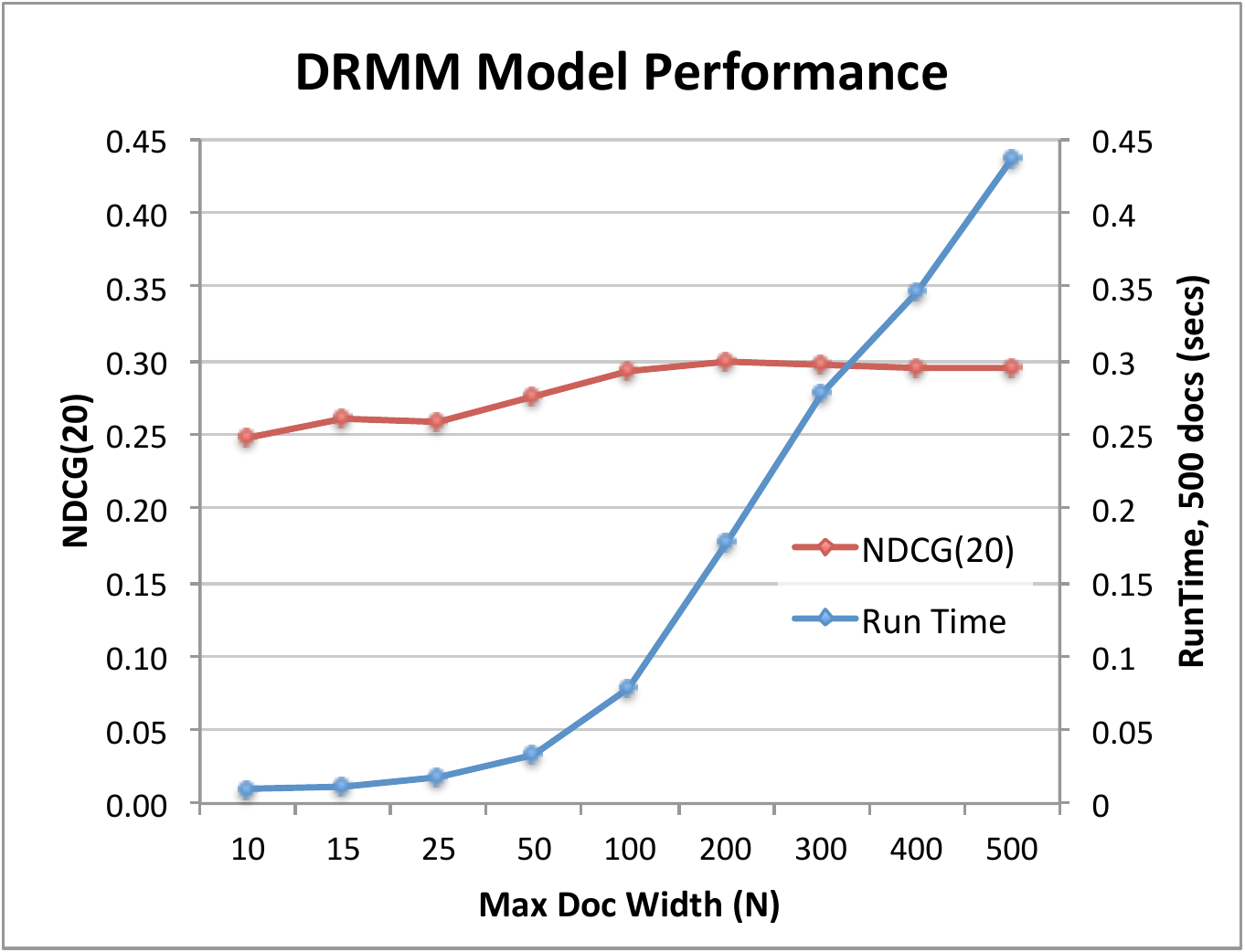}
	{\hspace{4.5em}(b)}
	\caption{(a) Distribution of Document Text widths. \ (b)~DRMM performance by max document width $N$.}
	\label{fig:DocWidthsAndDrmmPerf}
\end{figure}

\subsubsection{The DRMM Model.}

The Deep Relevance Matching Model (DRMM) is a recent example of the Local Interaction approach to text relevance, described in \cite{Guo-et-al:2016} to outperform several previous neural models on the Robust04 and ClueWeb-09-Cat-B datasets. While it is a simple model with only 162 trainable parameters, it begins by computing the cosine similarity between the embeddings of each document and query word pair, which dominates the model's computational cost. We implemented the DRMM model as described in \cite{Guo-et-al:2016}, using the Krovetz word stemmer during text tokenization, stopwords removed from queries, and the CBOW method of \cite{Mikolov:2013:NIPS} to compute word embeddings.

We tested DRMM on increasing values of $N$ (maximum document width) and found the ranking metrics stopped improving after a width of 200 words (figure~\ref{fig:DocWidthsAndDrmmPerf}b). DRMM uses the same pairwise loss function; we found different sample-weighting schemes to have an insignificant effect on the metrics. We report metrics for the version using square-root weighting and $N=200$.

\subsection{Metrics}

Each of the following metrics has values in the range $[0, 1]$, with higher values for better rankings. Scoring ties in all the compared approaches were resolved by sorting on decreasing document-id.

\subsubsection{NDCG}

Discounted Cumulative Gain (DCG) \cite{Jarvelin:2000} is a relevance and rank correlation metric that penalizes placement of relevant documents at lower ranks, computed as:
\[
DCG(n) = \sum_{i=1}^{n} \frac{2^{rel(i)} - 1}{\log_2(i + 1)}
\]
where $n$ is the rank to which DCG is accumulated, and $rel(i) \ge 0$ is the relevance level of the document placed at rank $i$. Normalized Discounted Cumulative Gain (NDCG) then measures the relative DCG of a ranking compared to the best possible ranking for that data: $NDCG(n) = DCG(n) / IDCG(n)$, where IDCG(n) is the DCG(n) for the ideal ranking. When there are multiple queries, NDCG refers to the mean value across queries. We use the scaled relevance levels (section~\ref{sec:TheData}), and quote ``NDCG.20'' metrics for $n=20$.

\subsubsection{Precision at Rank and MAP}

Average Precision \cite{Buckley:2000} measures, for a single query, the precision observed in a ranking up to the rank of each relevant document, averaged over the number of relevant documents for that query. It is thus a ranking measure that factors out the size of the ranked list and the number of relevant documents, without any rank-based penalization or discounting. We quote the Mean Average Precision (MAP), which is the mean of the Average Precision across queries in our test dataset. We also quote some Precision at rank $n$ metrics (``Prec.$n$'' in the tables).

\section{Evaluation}

\begin{table}
\small
\begin{center}
\begin{tabular}{$l^l^l^l^l^l}
\toprule
{} & NDCG.20 & MAP & Prec.5  \\
\midrule
rev DocID	          & 0.141	& 0.455	& 0.344		\\
BM25-Title$^1$	 & 0.325	& 0.567	& 0.591	 \\
UQLM-Title	        & 0.314	& 0.560	& 0.574	 \\
WMD-Title$^2$	& 0.329$^{=1}$	& 0.579$^{+1}$	& 0.603$^{+1}$	 \\
\midrule
DRMM	               & 0.300$^{-1}$	& 0.545$^{-1}$	& 0.549$^{-1}$	 \\
SevMos-C1$^3$	& 0.352$^{+2}$	& 0.597$^{+2}$	& 0.625$^{+2}$	 \\
SevMos-C3$^4$	& 0.373$^{+3}$	& 0.594$^{=3,+2}$	& 0.626$^{=3,+2}$ \\
\midrule
Delta-32$^5$	   & 0.365$^{+3,-4}$	& 0.601$^{+3,4}$	& 0.634$^{+3,4}$	 \\
Delta-32-Lex3	\rowstyle{\bfseries}  & 0.394$^{+4,5}$	& 0.609$^{+4,5}$	& 0.646$^{+4,5}$ \\
\bottomrule
\end{tabular}
\end{center}
\caption{\label{tab:Metrics.Full}Ranking metrics on the Full test data. \textmd{The superscripts indicate statistical comparisons: `+' for increase, `-' for decrease, `=' for equivalent, to a 99\% confidence using a paired t-test. The comparison baselines are indicated with numbers 1 through 5, as marked in the first column. Highest values are in bold.}}
\end{table}

\subsection{Test Metrics}

We compare ranking performance of two versions of the Delta model against the other approaches. The `Delta-32' model uses $n_f=32$ feature maps in the Convolutional stage, and no Lexical Match features. The `Delta-32-Lex3' version of the model adds the following three Lexical Match features: BM25 on the Document Abstract, IDF weighted Jaccard Similarity between the Query and the Document Title, and IDF-weighted proportion of unique Query words in the Document Title. These features were selected using greedy search on the list of 18 described earlier, with NDCG.20 on the validation data as the selection criterion. The feature selection was limited to three to control model run-time cost.

We begin with ranking metrics for the various approaches on the full test data (table~\ref{tab:Metrics.Full}). The first row in the table provides metrics for an uninformed ranker, where documents are ranked on decreasing document id, to provide a low threshold of performance. The table indicates whether there was a statistically significant change (to a 99\% confidence, using a paired t-test) against a baseline. Among the non-trained relevance models, Word Mover's Distance (WMD) performed at least as well as BM25, with no change in NDCG.20, but an improvement in MAP and Prec.5 (Precision at rank 5), while the Query Language Model (UQLM) did not match BM25's level of performance. Among the trained neural models, DRMM performed the worst, with lower metrics than even BM25. The SevMos-C1 model performed better overall than WMD, and SevMos-C3 further improved the NDCG.20 score.

Among the Delta models, Delta-32 showed better performance overall than SevMos-C1. However compared to SevMos-C3, its NDCG.20 score was lower, while the MAP and Prec.5 scores were higher. The Delta-32-Lex3 model exhibited the best metrics overall, bettering both Delta-32 and SevMos-C3. These gains were observed not just in the relevance-weighted NDCG.20 metrics that use our derived scaled relevance levels, but also in the MAP and Prec.5 precision-based metrics that use a binary notion of relevance.

The good performance of SevMos-C3 over SevMos-C1 demonstrates the benefits of using a convolutional stack. Combining these elements with the Delta matrix implementation of the Local Interaction architecture yields even better results, as depicted in the metrics for Delta-32-Lex3.

\begin{table}
\small
\begin{center}
\begin{tabular}{$l^l^l^l^l^l}
\toprule
{} & NDCG.20 & MAP & Prec.5  \\
\midrule
rev DocID           & 0.081 & 0.413 & 0.310  \\
BM25-Title$^1$ & 0.233 & 0.474 & 0.490  \\
WMD-Title$^2$  & 0.243$^{+1}$ & 0.483$^{+1}$ & 0.496$^{=1}$  \\
\midrule
DRMM                 & 0.242$^{+1,=2}$ & 0.461$^{-1,2}$ & 0.462$^{-1,2}$  \\
SevMos-C1$^3$  & 0.290$^{+2}$ & 0.510$^{+2}$ & 0.538$^{+2}$  \\
SevMos-C3$^4$  & 0.304$^{+2,3}$ & 0.502$^{+2,-3}$ & 0.535$^{+2,=3}$  \\
\midrule
Delta-32$^5$     & 0.296$^{+2,=3}$ & 0.513$^{+2,=3}$ & 0.550$^{+2,3}$  \\
Delta-32-Lex3 \rowstyle{\bfseries}    & 0.326$^{+4,5}$ & 0.522$^{+4,5}$ & 0.560$^{+4,5}$  \\
\bottomrule
\end{tabular}
\end{center}
\caption{\label{tab:Metrics.Neg20plus}Ranking metrics on the `Neg20+' test data}
\end{table}

\begin{table}
\small
\begin{center}
\begin{tabular}{$l^l^l^l^l^l}
\toprule
{} & NDCG.20 & MAP & Prec.5  \\
\midrule
rev DocID & 0.191 & 0.488 & 0.364  \\
BM25-Title$^1$ & 0.333 & 0.593 & 0.604  \\
WMD-Title$^2$ & 0.330$^{=1}$ & 0.603$^{+1}$ & 0.614$^{+1}$  \\
\midrule
DRMM & 0.318$^{-1}$ & 0.580$^{-1}$ & 0.578$^{-1}$  \\
SevMos-C1$^3$ & 0.358$^{+2}$ & 0.624$^{+2}$ & 0.642$^{+2}$  \\
SevMos-C3$^4$ & 0.375$^{+2,3}$ & 0.621$^{+2,=3}$ & 0.640$^{+2,=3}$  \\
\midrule
Delta-32$^5$ & 0.382$^{+3}$ & 0.629$^{=3}$ & 0.648$^{=3}$  \\
Delta-32-Lex3 \rowstyle{\bfseries}  & 0.413$^{+4,5}$ & 0.638$^{+4,5}$ & 0.666$^{+4,5}$  \\
\bottomrule
\end{tabular}
\end{center}
\caption{\label{tab:Metrics.OneNewWord}Ranking metrics on the `OneNewWord' test data. \textmd{The significance for DRMM's NDCG.20 comparison is to a confidence of 95\%.}}
\end{table}

To evaluate performance on the {\em Under-Specified Query Problem}, we compared metrics on the `Neg20+' subset of the test data (table~\ref{tab:Metrics.Neg20plus}). These are harder queries to rank for, since many non-relevant documents contain all the query words. Not surprisingly, the scores for all the models dropped. WMD was still the benchmark among untrained models, and DRMM the lowest performing deep learning model, although it did have better NDCG.20 score than BM25. The Delta-32-Lex3 model again exhibited the best overall performance.

To evaluate model robustness, we looked at performance on the `OneNewWord' and `AllNewWords' subsets of the test data (tables~\ref{tab:Metrics.OneNewWord} and \ref{tab:Metrics.AllNewWords}). The general trend among models here was the same as for the full test data, with DRMM performing no better than BM25, and SevMos-C3 better than SevMos-C1 and WMD. The Delta-32-Lex3 model showed the best overall performance, demonstrating that it was the most robust approach among the models tested. As a final note, some of the models exhibited better metrics on these subsets than the overall data because they tend to do better on shorter queries, and these test subsets had a higher concentration of shorter queries.

\begin{table}
\small
\begin{center}
\begin{tabular}{$l^l^l^l}
\toprule
{} & NDCG.20 & MAP & Prec.5  \\
\midrule
rev DocID           & 0.195 & 0.508 & 0.389  \\
BM25-Title$^1$ & 0.309 & 0.581 & 0.586  \\
WMD-Title$^2$  & 0.306$^{=1}$ & 0.590$^{+1}$ & 0.595$^{=1}$  \\
\midrule
DRMM                 & 0.311$^{=1,2}$ & 0.578$^{=1,-2}$ & 0.570$^{=1,-2}$  \\
SevMos-C1$^3$ & 0.344$^{+2}$ & 0.615$^{+2}$ & 0.624$^{+2}$  \\
SevMos-C3$^4$ & 0.355$^{+2,=3}$ & 0.614$^{+2,=3}$ & 0.632$^{+2,=3}$  \\
\midrule
Delta-32$^5$     & 0.362$^{+3,=4}$ & 0.622$^{+3,4}$ & 0.638$^{=3,4}$  \\
Delta-32-Lex3 \rowstyle{\bfseries}  & 0.400$^{+4,5}$ & 0.633$^{+4,5}$ & 0.661$^{+4,5}$  \\
\bottomrule
\end{tabular}
\end{center}
\caption{\label{tab:Metrics.AllNewWords}Ranking metrics on the `AllNewWords' test data}
\end{table}

\subsection{Impact of Different Features}

\begin{figure}
	\includegraphics[width=\columnwidth]{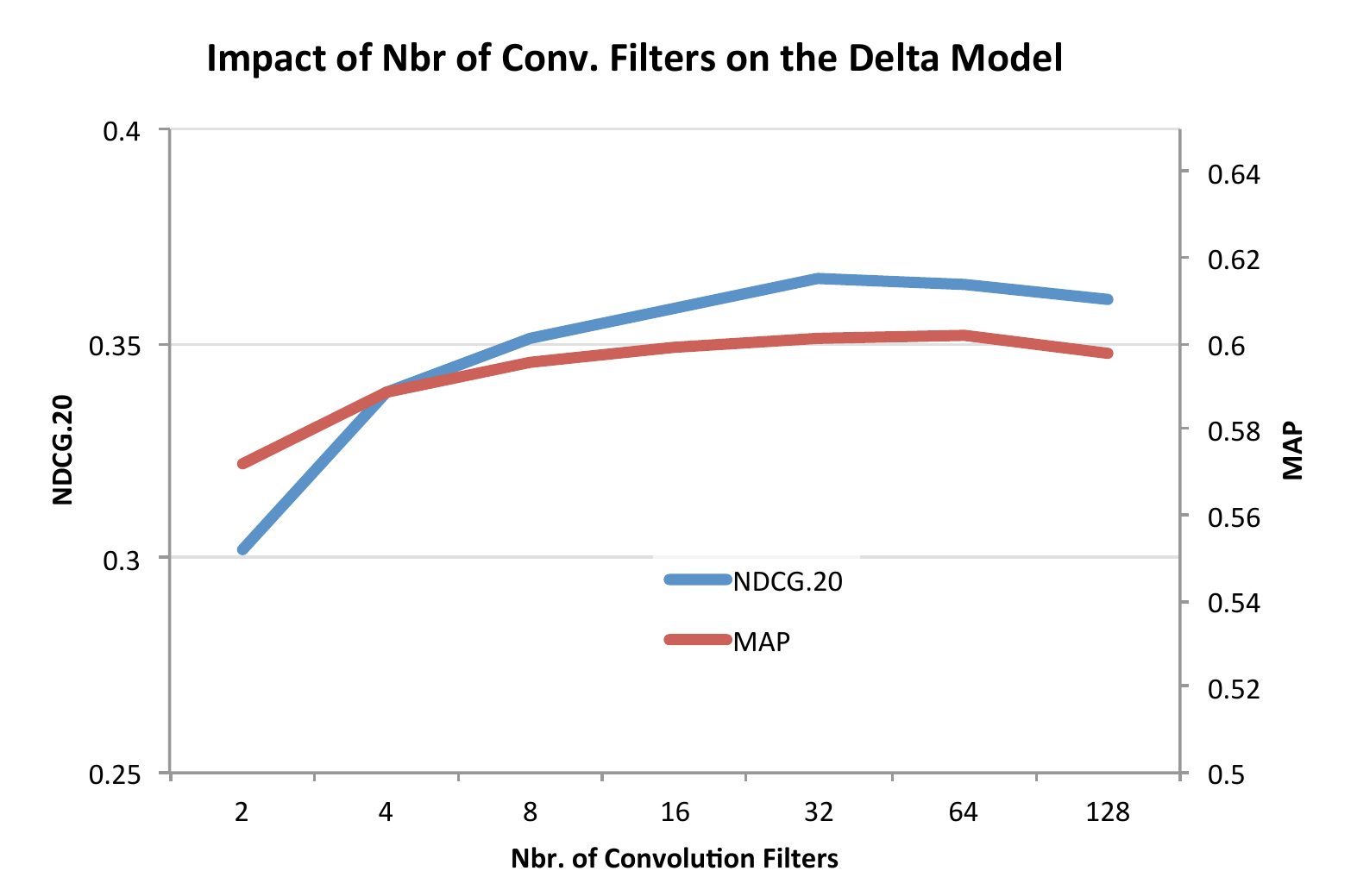}
	\caption{Comparing the impact of number of filters $n_f$ (feature maps) in the Convolutional Stage of the Delta Model.}
	\label{fig:NbrConvFilters}
\end{figure}

Next we look at how different aspects of the Delta model affected its performance. The Convolutional stage extracts match-related features from word $n$-grams in the document. The parameter $n_f$ controls the number of such features, and its effect on ranking is charted in figure~\ref{fig:NbrConvFilters}. Both NDCG.20 and MAP improved as $n_f$ increased till around 32 filters, and then performance leveled off and then dropped slightly. At larger number of filters, the model becomes more complex, but this increase in complexity does not continue to yield better performance. More complex models are more likely to overfit the training data, and perhaps learning rate decay techniques might help converge to a better solution, an area to be explored further. However since our goal was to construct a fast model for use in a search engine, we had a preference for smaller models, and $n_f=32$ provided a good balance between speed and performance.

\begin{table}
\begin{center}
\begin{tabular}{$l^l^l}
\toprule
{} & NDCG.20 & MAP  \\
\midrule
Delta-32, no Difference vectors & 0.323 & 0.574 \\
Delta-32, no Delta features & 0.333 & 0.584 \\
Delta-32 & 0.365 & 0.601 \\
Delta-32-Lex3, (top 3 Lex features) & 0.394 & 0.609 \\
\bottomrule
\end{tabular}
\end{center}
\caption{\label{tab:Metrics.Features}Comparing the impact of different features on the ranking metrics for the Delta model}
\end{table}

Table~\ref{tab:Metrics.Features} compares four different versions of the Delta model with $n_f=32$. The Delta Stage of the model computes, for each document word, a difference vector against the closest query word, and three `Delta features': the cosine similarity, euclidean distance, and normalized proximity. The first two rows of table~\ref{tab:Metrics.Features} show the impact of removing the difference vectors and the Delta features from the Delta-32 model, on the ranking metrics for the test data. Both show a significant drop in performance compared to Delta-32. Finally, as reviewed above, adding the 3 lexical match features resulted in a significant improvement for the Delta-32-Lex3 model over the Delta-32 model. 

In our greedy search for selecting the most useful lexical match features, BM25 on Document Abstract showed the most impact because it compensated for the truncation of the abstract when the document was limited to 50 words. The other two lexical match features are useful in accounting for the matches of out-of-vocabulary query terms (as discussed in \cite{Severyn:2015}), and also for providing for query term significance through IDF-weighting.

\subsection{Evaluation as a Reranker}

Our goal for this research was to develop a good textual relevance model whose output could be used as a factor in a reranker (which would also use other factors like document meta-data) in a search engine like PubMed. We set a target of re-ranking the top 500 documents as produced by a fast naive ranker, and a run-time performance constraint on the relevance model of scoring the 500 documents in under 0.1 seconds on a GPU, yielding a throughput of at least 10 queries per second per GPU. While the complete design of a two-round ranker was outside the scope of this project, we compared the candidate relevance models on the (up to) top 500 results for the same test queries, as ranked by PubMed's implementation of BM25, which also incorporates query expansion terms.

\subsubsection{Ranking Metrics.}
Table \ref{tab:Metrics.Top500} shows the ranking metrics for our candidate and baseline models on the top 500 results test data, with models sorted on the NDCG.20 metric. This data provided a particular challenge for relevance models, since on average less than 4\% of the documents were relevant to the corresponding query. It also did not have the same selection bias present in the training data which was extracted from clicks on results sorted by PubMed's relevance ranker, a more complex expression that includes BM25 as one of the factors. As a result, all the metrics were lower than for the previous test data. However the general trends among the models was the same, with Delta-32-Lex3 the best model overall by a large margin, followed by SevMos-C3. An area worth exploring further is whether adding some randomly sampled documents from deep in the search results (e.g. \cite{Joachims:2002b}) could help overcome some of this selection bias.

\begin{table}
\begin{center}
\begin{tabular}{$l^l^l}
\toprule
{} & NDCG.20 & MAP  \\
\midrule
Original sort order & 0.025 & 0.109 \\
BM25-Title & 0.032 & 0.077 \\
SevMos-C1 & 0.083 & 0.104 \\
DRMM & 0.106 & 0.135 \\
WMD-Title & 0.122 & 0.153 \\
Delta-32 & 0.141 & 0.154 \\
SevMos-C3 & 0.160 & 0.151 \\
Delta-32-Lex3 \rowstyle{\bfseries}  & 0.191 & 0.188 \\
\bottomrule
\end{tabular}
\end{center}
\caption{\label{tab:Metrics.Top500}Ranking metrics for re-ranking the top 500 results as provided by PubMed's BM25, sorted on NDCG.20}
\end{table}

\subsubsection{Run-time Cost.}

As discussed above, the Independent Semantic Vector approach like that used in the SevMos models \cite{Severyn:2015} is particularly attractive for use in search engines, because the document semantic vectors (e.g. pooled output from the convolution stage in SevMos) can be pre-computed and stored, the query vector needs to be computed just once per search, and the remaining computation (the similarity measure and classifier stage in SevMos) is fairly small and fast. This caching is not possible in the Local Interaction approaches like the DRMM and Delta models. In these two models, the computation is dominated by the cost to compare each pair of document and query words, so we look to reducing the size of the document by truncation to control the computation. In scientific literature, authors are strongly motivated to provide a highly informative and noise-free document Title and Abstract, making it particularly amenable to this approach.

We measured the time each model took to score 500 documents for a single query on a NVIDIA GeForce GTX TITAN X GPU, including the time to transfer the data to the GPU in a single batch, but not including the time to load the model and its parameters (including embeddings). While best ranking metrics for DRMM were obtained for $N=200$ (with NDCG.20 = 0.300), its run-time of 0.177 seconds, corresponding to a throughput of 5.6 queries per second per GPU, exceeded our criterion; DRMM for $N=100$ came in at 0.079 seconds (throughput 12.7 qps per GPU), but a slightly lower overall NDCG.20 of 0.293 (figure~\ref{fig:DocWidthsAndDrmmPerf}b). The `Delta-32-Lex3' model had a run-time of 0.049 secs, (20 qps per GPU). As a comparison the full `SevMos-C3' model scored 500 documents in 0.040 seconds (25 qps per GPU).

\subsection{Example Queries}

We compare the rankings of the `Delta-32-Lex3', `SevMos-C3' and `WMD' relevance models on some interesting queries. The NDCG at 20 at that query is quoted for each model, along with Titles and relevance levels for the top 3 scoring documents, and the relevance leves for the next 4 documents. The examples demonstrate the Delta model's ability to detect relevance in documents without exact match of query terms (addressing the {\em term mismatch problem}), and where the context of the match is also important.

\subsubsection{Query: \texttt{countermovement jump}}
The word {\em countermovement} did not occur in training or validation queries. This is also an example where relevance depends on the other words in the document text besides those matching the query. Number of documents in the test dataset: relevant = 17, non-relevant = 16. Top relevance levels: 21.5, 18.1, 11.2, 7.8, 4.4.


As ranked by \textbf{Delta-32-Lex3} (NDCG.20 = 0.98):
\begin{enumerate}[topsep=0pt,itemsep=0pt,parsep=1pt]
\renewcommand{\labelenumi}{\roman{enumi}.}\footnotesize 
\item (21.5)  Determinants of countermovement jump performance: a kinetic and kinematic analysis.

\item (11.2)  Which drop jump technique is most effective at enhancing countermovement jump ability, ``countermovement'' drop jump or ``bounce'' drop jump?

\item (4.4)  The MARS for squat, countermovement, and standing long jump performance analyses: are measures reproducible?

\item[] \hspace{-1em} Next four relevance levels = 4.4, 0, 18.1, 4.4.
\end{enumerate}

As ranked by \textbf{SevMos-C3} (NDCG.20 = 0.32):
\begin{enumerate}[topsep=0pt,itemsep=0pt,parsep=1pt]
\renewcommand{\labelenumi}{\roman{enumi}.}\footnotesize
\item (11.2)  Which drop jump technique is most effective at enhancing countermovement jump ability, ``countermovement'' drop jump or ``bounce'' drop jump?

\item (0)  A mechanics comparison between landing from a countermovement jump and landing from stepping off a box.  

\item (4.4)  Comparison of acute countermovement jump responses after functional isometric and dynamic half squats.

\item[] \hspace{-1em} Next four relevance levels = 0, 4.4, 0, 21.5.
\end{enumerate}

As ranked by \textbf{WMD} (NDCG.20 = 0.5):
\begin{enumerate}[topsep=0pt,itemsep=0pt,parsep=1pt]
\renewcommand{\labelenumi}{\roman{enumi}.}\footnotesize
\item (11.2)  Which drop jump technique is most effective at enhancing countermovement jump ability, ``countermovement'' drop jump or ``bounce'' drop jump?

\item (0)  Reductions in Sprint Paddling Ability and Countermovement Jump Performance After Surfing Training.

\item (21.5)  Determinants of countermovement jump performance: a kinetic and kinematic analysis.

\item[] \hspace{-1em} Next four relevance levels = 4.4, 0, 0, 0.
\end{enumerate}

\subsubsection{Query: \texttt{oesophageal cancer review}}
The word {\em oesopha-geal} did not occur in training or validation queries. The word {\em review} does not occur in the title of all relevant documents. The three models successfully located alternative spellings of the word. Number of documents in the test dataset: relevant = 22, non-relevant = 28. Top relevance levels: 11.7, and four at 7.1.


As ranked by \textbf{Delta-32-Lex3} (NDCG.20 = 0.94):
\begin{enumerate}[topsep=0pt,itemsep=0pt,parsep=1pt]
\renewcommand{\labelenumi}{\roman{enumi}.}\footnotesize
\item (11.7)  Esophageal cancer: Recent advances in screening, targeted therapy, and management.

\item (0)  Outcomes in the management of esophageal cancer.

\item (0)  Current advances in esophageal cancer proteomics.

\item[] \hspace{-1em} Next four relevance levels = 5.6, 5.6, 0, 7.1.
\end{enumerate}

As ranked by \textbf{SevMos-C3} (NDCG.20 = 0.05):
\begin{enumerate}[topsep=0pt,itemsep=0pt,parsep=1pt]
\renewcommand{\labelenumi}{\roman{enumi}.}\footnotesize
\item (5.6)  Esophageal Cancer Staging.

\item (2.5)  Esophageal cancer: staging system and guidelines for staging and treatment.

\item (0) Outcomes in the management of esophageal cancer.

\item[] \hspace{-1em} Next four relevance levels = 0, 0, 5.6, 0.
\end{enumerate}

As ranked by \textbf{WMD} (NDCG.20 = 0.05):
\begin{enumerate}[topsep=0pt,itemsep=0pt,parsep=1pt]
\renewcommand{\labelenumi}{\roman{enumi}.}\footnotesize
\item (5.6)  Esophageal Cancer Staging.

\item (7.1)  Endoscopic Management of Early Esophageal Cancer.

\item (0) Endoscopic treatment of early esophageal cancer.

\item[] \hspace{-1em} Next four relevance levels = 0, 0, 5.6, 0.
\end{enumerate}

\subsubsection{Query: \texttt{chronic headache and depression review}}
All three models were able to leverage word vectors to relate headache to migraine. Delta-32-Lex3 placed the most relevant document (``{\small Psychological Risk Factors in Headache}'', rel. level = 10) at rank 5. It did not feature in the top 10 of any of the other neural and lexical models tested. This example demonstrates the need for deeper semantic modeling, where the Delta model has some limited success. Number of documents in the test dataset: relevant = 23, non-relevant = 18. Top relevance levels: a 10, and four at 5.5.


\begin{samepage}
As ranked by \textbf{Delta-32-Lex3} (NDCG.20 = 0.4):
\begin{enumerate}[topsep=0pt,itemsep=0pt,parsep=1pt]
\renewcommand{\labelenumi}{\roman{enumi}.}\footnotesize
\item (5.5)  Comprehensive management of headache and depression.

\item (0)  Medication overuse headache.

\item (0)  Clinical features and mechanisms of chronic migraine and medication-overuse headache.

\item[] \hspace{-1em} Next four relevance levels = 0, 10, 5.5, 5.5.
\end{enumerate}
\end{samepage}

As ranked by \textbf{SevMos-C3} (NDCG.20 = 0.3):
\begin{enumerate}[topsep=0pt,itemsep=0pt,parsep=1pt]
\renewcommand{\labelenumi}{\roman{enumi}.}\footnotesize 
\item (5.5)  Chronic headaches and the neurobiology of somatization.

\item (2.7)  Pathophysiology of migraine -- from molecular to personalized medicine.

\item (0)  Medication overuse headache.

\item[] \hspace{-1em} Next four relevance levels = 5.5, 2.5, 5.5, 0.
\end{enumerate}

As ranked by \textbf{WMD} (NDCG.20 = 0.35):
\begin{enumerate}[topsep=0pt,itemsep=0pt,parsep=1pt]
\renewcommand{\labelenumi}{\roman{enumi}.}\footnotesize 
\item (5.5)  Comprehensive management of headache and depression.

\item (5.5)  Migraine and depression: biological aspects.

\item (5.5) Migraine and depression.

\item[] \hspace{-1em} Next four relevance levels = 5.5, 0, 0, 0.
\end{enumerate}

\section{Conclusion}

While deep learning models for text understanding have made dramatic gains in recent years, they have tended to be large and slow. The challenge in information retrieval is still one of combining predictive power with low run-time overhead. This is also true when the corpus is scientific literature. 

We described the Delta Relevance model, a new deep learning model for text relevance, targeted for information retrieval in biomedical science literature. The main innovation in the model is to base the modeling function on differences and distances between word embeddings, as captured in the Delta features, rather than directly using the embeddings themselves like most NLP approaches. Other researchers have shown the benefit of training word embeddings as part of the model. That was not feasible here since the training data had only 20k queries compared to a vocabulary size of 200k. Using Delta features as the input to the model helped adapt the pre-trained word embeddings to our task.

To achieve our goal of fast run-time, we used a convolutional network rather than the recurrent approaches popular in most neural NLP models. Using a stack of narrow convolutional layers instead of a single wide convolution gave the model more power.

We showed that the Delta model outperformed comparable recent approaches in ranking metrics when trained and evaluated on data derived from the PubMed search engine click logs. We demonstrated that the model was robust, despite being trained on a relatively small amount of data, and the model was fast enough for use in an on-line search engine.

The Delta model might be especially suited to scientific literature in taking advantage of the high quality of the document Title and first few sentences of the Abstract.  We believe that the previously observed good performance of the DRMM approach was on documents that were quite noisy (they contain a lot of meta data in the document text). 
An area worth exploring further, for its potential in improving both prediction performance and run-time costs, is pre-processing a document to extract significant portions for evaluating relevance, thus reducing the size of the input at run-time. Another area to explore is the benefit from retaining sentence and grammatical structure in document text.

\begin{acks}
This research was supported by the Intramural Research Program of the NIH, National Library of Medicine.
\end{acks}

%% file: delta_paper.bbl

\begin{thebibliography}{40}


\ifx \showCODEN    \undefined \def \showCODEN     #1{\unskip}     \fi
\ifx \showDOI      \undefined \def \showDOI       #1{#1}\fi
\ifx \showISBNx    \undefined \def \showISBNx     #1{\unskip}     \fi
\ifx \showISBNxiii \undefined \def \showISBNxiii  #1{\unskip}     \fi
\ifx \showISSN     \undefined \def \showISSN      #1{\unskip}     \fi
\ifx \showLCCN     \undefined \def \showLCCN      #1{\unskip}     \fi
\ifx \shownote     \undefined \def \shownote      #1{#1}          \fi
\ifx \showarticletitle \undefined \def \showarticletitle #1{#1}   \fi
\ifx \showURL      \undefined \def \showURL       {\relax}        \fi
\providecommand\bibfield[2]{#2}
\providecommand\bibinfo[2]{#2}
\providecommand\natexlab[1]{#1}
\providecommand\showeprint[2][]{arXiv:#2}

\bibitem[\protect\citeauthoryear{Bodenreider}{Bodenreider}{2004}]%
        {Bodenreider:2004}
\bibfield{author}{\bibinfo{person}{Olivier Bodenreider}.}
  \bibinfo{year}{2004}\natexlab{}.
\newblock \showarticletitle{The Unified Medical Language System (UMLS):
  integrating biomedical terminology}.
\newblock   \bibinfo{volume}{32 (Database issue)} (\bibinfo{year}{2004}),
  \bibinfo{pages}{D267--D270}.
\newblock


\bibitem[\protect\citeauthoryear{Buckley and Voorhees}{Buckley and
  Voorhees}{2000}]%
        {Buckley:2000}
\bibfield{author}{\bibinfo{person}{Chris Buckley} {and}
  \bibinfo{person}{Ellen~M. Voorhees}.} \bibinfo{year}{2000}\natexlab{}.
\newblock \showarticletitle{Evaluating Evaluation Measure Stability}. In
  \bibinfo{booktitle}{\emph{Proceedings of the 23rd Annual International ACM
  SIGIR Conference on Research and Development in Information Retrieval}}
  \emph{(\bibinfo{series}{SIGIR '00})}. \bibinfo{publisher}{ACM},
  \bibinfo{address}{New York, NY, USA}, \bibinfo{pages}{33--40}.
\newblock
\showISBNx{1-58113-226-3}


\bibitem[\protect\citeauthoryear{Chiu, Crichton, Korhonen, and Pyysalo}{Chiu
  et~al\mbox{.}}{2016}]%
        {Chiu-etal:2016:BioNLP}
\bibfield{author}{\bibinfo{person}{Billy Chiu}, \bibinfo{person}{Gamal
  Crichton}, \bibinfo{person}{Anna Korhonen}, {and} \bibinfo{person}{Sampo
  Pyysalo}.} \bibinfo{year}{2016}\natexlab{}.
\newblock \showarticletitle{How to Train good Word Embeddings for Biomedical
  NLP}. In \bibinfo{booktitle}{\emph{Proceedings of the 15th Workshop on
  Biomedical Natural Language Processing}} \emph{(\bibinfo{series}{BioNLP
  '16})}. \bibinfo{publisher}{ACL}, \bibinfo{pages}{166--174}.
\newblock


\bibitem[\protect\citeauthoryear{Diaz, Mitra, and Craswell}{Diaz
  et~al\mbox{.}}{2016}]%
        {Diaz-etal:2016}
\bibfield{author}{\bibinfo{person}{Fernando Diaz}, \bibinfo{person}{Bhaskar
  Mitra}, {and} \bibinfo{person}{Nick Craswell}.}
  \bibinfo{year}{2016}\natexlab{}.
\newblock \showarticletitle{Query Expansion with Locally-Trained Word
  Embeddings}. In \bibinfo{booktitle}{\emph{Proceedings of ACL 2016}}.
  \bibinfo{publisher}{ACL}, \bibinfo{address}{Berlin, Germany},
  \bibinfo{pages}{367--377}.
\newblock


\bibitem[\protect\citeauthoryear{Dogan, Murray, N{\'e}v{\'e}ol, and Lu}{Dogan
  et~al\mbox{.}}{2009}]%
        {Dogan:2009}
\bibfield{author}{\bibinfo{person}{Rezarta~Islamaj Dogan},
  \bibinfo{person}{G.~Craig Murray}, \bibinfo{person}{Aur{\'e}lie
  N{\'e}v{\'e}ol}, {and} \bibinfo{person}{Zhiyong Lu}.}
  \bibinfo{year}{2009}\natexlab{}.
\newblock \showarticletitle{Understanding
  {PubMed}{\textsuperscript{\textregistered}} user search behavior through log
  analysis}.
\newblock \bibinfo{journal}{\emph{Database}}  \bibinfo{volume}{2009}
  (\bibinfo{year}{2009}), \bibinfo{pages}{bap018}.
\newblock


\bibitem[\protect\citeauthoryear{Duchi, Hazan, and Singer}{Duchi
  et~al\mbox{.}}{2011}]%
        {Duchi:2011}
\bibfield{author}{\bibinfo{person}{John Duchi}, \bibinfo{person}{Elad Hazan},
  {and} \bibinfo{person}{Yoram Singer}.} \bibinfo{year}{2011}\natexlab{}.
\newblock \showarticletitle{Adaptive subgradient methods for online learning
  and stochastic optimization}.
\newblock \bibinfo{journal}{\emph{Journal of Machine Learning Research}}
  \bibinfo{volume}{12} (\bibinfo{date}{Jul} \bibinfo{year}{2011}),
  \bibinfo{pages}{2121--2159}.
\newblock


\bibitem[\protect\citeauthoryear{Durao, Bayyapu, Xu, Dolog, and Lage}{Durao
  et~al\mbox{.}}{2013}]%
        {Durao-etal:2013}
\bibfield{author}{\bibinfo{person}{Frederico Durao}, \bibinfo{person}{Karunakar
  Bayyapu}, \bibinfo{person}{Guandong Xu}, \bibinfo{person}{Peter Dolog}, {and}
  \bibinfo{person}{Ricardo Lage}.} \bibinfo{year}{2013}\natexlab{}.
\newblock \bibinfo{booktitle}{\emph{Medical Information Retrieval Enhanced with
  User's Query Expanded with Tag-Neighbors}}.
\newblock \bibinfo{publisher}{Springer New York}, \bibinfo{address}{New York,
  NY}, \bibinfo{pages}{17--40}.
\newblock
\showISBNx{978-1-4614-8495-0}
\urldef\tempurl%
\url{https://doi.org/10.1007/978-1-4614-8495-0_2}
\showDOI{\tempurl}


\bibitem[\protect\citeauthoryear{Furnas, Landauer, Gomez, and Dumais}{Furnas
  et~al\mbox{.}}{1987}]%
        {Furnas-et-al:1987}
\bibfield{author}{\bibinfo{person}{G.~W. Furnas}, \bibinfo{person}{T.~K.
  Landauer}, \bibinfo{person}{L.~M. Gomez}, {and} \bibinfo{person}{S.~T.
  Dumais}.} \bibinfo{year}{1987}\natexlab{}.
\newblock \showarticletitle{The Vocabulary Problem in Human system
  Communication}.
\newblock \bibinfo{journal}{\emph{CACM}} \bibinfo{volume}{30},
  \bibinfo{number}{11} (\bibinfo{date}{nov} \bibinfo{year}{1987}),
  \bibinfo{pages}{964--971}.
\newblock
\showISSN{0001-0782}


\bibitem[\protect\citeauthoryear{Gao, Pantel, Gamon, He, and Deng}{Gao
  et~al\mbox{.}}{2014}]%
        {Gao-et-al:2014:EMNLP}
\bibfield{author}{\bibinfo{person}{Jianfeng Gao}, \bibinfo{person}{Patrick
  Pantel}, \bibinfo{person}{Michael Gamon}, \bibinfo{person}{Xiaodong He},
  {and} \bibinfo{person}{Li Deng}.} \bibinfo{year}{2014}\natexlab{}.
\newblock \showarticletitle{Modeling Interestingness with Deep Neural
  Networks}. In \bibinfo{booktitle}{\emph{Proceedings of EMNLP 2014}}.
  \bibinfo{publisher}{ACL}, \bibinfo{address}{Doha, Qatar},
  \bibinfo{pages}{2--13}.
\newblock


\bibitem[\protect\citeauthoryear{Goodfellow, Bengio, and Courville}{Goodfellow
  et~al\mbox{.}}{2016}]%
        {Goodfellow-et-al:2016}
\bibfield{author}{\bibinfo{person}{Ian Goodfellow}, \bibinfo{person}{Yoshua
  Bengio}, {and} \bibinfo{person}{Aaron Courville}.}
  \bibinfo{year}{2016}\natexlab{}.
\newblock \bibinfo{booktitle}{\emph{Deep Learning}}.
\newblock \bibinfo{publisher}{MIT Press}, \bibinfo{address}{Cambridge, MA}.
\newblock
\newblock
\shownote{\url{http://www.deeplearningbook.org}.}


\bibitem[\protect\citeauthoryear{Guo, Fan, Ai, and Croft}{Guo
  et~al\mbox{.}}{2016}]%
        {Guo-et-al:2016}
\bibfield{author}{\bibinfo{person}{Jiafeng Guo}, \bibinfo{person}{Yixing Fan},
  \bibinfo{person}{Qingyao Ai}, {and} \bibinfo{person}{W.~Bruce Croft}.}
  \bibinfo{year}{2016}\natexlab{}.
\newblock \showarticletitle{A Deep Relevance Matching Model for Ad-hoc
  Retrieval}. In \bibinfo{booktitle}{\emph{Proceedings of CIKM 2016}}.
  \bibinfo{publisher}{ACM}, \bibinfo{address}{New York, NY, USA},
  \bibinfo{pages}{55--64}.
\newblock
\showISBNx{978-1-4503-4073-1}


\bibitem[\protect\citeauthoryear{Hersh, Price, and Donohoe}{Hersh
  et~al\mbox{.}}{2000}]%
        {Hersh-etal:2000}
\bibfield{author}{\bibinfo{person}{W. Hersh}, \bibinfo{person}{S. Price}, {and}
  \bibinfo{person}{L. Donohoe}.} \bibinfo{year}{2000}\natexlab{}.
\newblock \showarticletitle{Assessing Thesaurus-Based Query Expansion Using the
  UMLS Metathesaurus}. In \bibinfo{booktitle}{\emph{Proceedings of the AMIA
  Symposium}}. \bibinfo{pages}{344--348}.
\newblock


\bibitem[\protect\citeauthoryear{Hu, Lu, Li, and Chen}{Hu
  et~al\mbox{.}}{2014}]%
        {Hu-et-al:2014:NIPS}
\bibfield{author}{\bibinfo{person}{Baotian Hu}, \bibinfo{person}{Zhengdong Lu},
  \bibinfo{person}{Hang Li}, {and} \bibinfo{person}{Qingcai Chen}.}
  \bibinfo{year}{2014}\natexlab{}.
\newblock \showarticletitle{Convolutional Neural Network Architectures for
  Matching Natural Language Sentences}.
\newblock In \bibinfo{booktitle}{\emph{Advances in NIPS 27}},
  \bibfield{editor}{\bibinfo{person}{Z.~Ghahramani},
  \bibinfo{person}{M.~Welling}, \bibinfo{person}{C.~Cortes},
  \bibinfo{person}{N.~D. Lawrence}, {and} \bibinfo{person}{K.~Q. Weinberger}}
  (Eds.). \bibinfo{pages}{2042--2050}.
\newblock


\bibitem[\protect\citeauthoryear{Huang, He, Gao, Deng, Acero, and Heck}{Huang
  et~al\mbox{.}}{2013}]%
        {Huang:2013}
\bibfield{author}{\bibinfo{person}{Po-Sen Huang}, \bibinfo{person}{Xiaodong
  He}, \bibinfo{person}{Jianfeng Gao}, \bibinfo{person}{Li Deng},
  \bibinfo{person}{Alex Acero}, {and} \bibinfo{person}{Larry Heck}.}
  \bibinfo{year}{2013}\natexlab{}.
\newblock \showarticletitle{Learning deep structured semantic models for web
  search using clickthrough data}. In \bibinfo{booktitle}{\emph{Proceedings of
  CIKM 2013}}. \bibinfo{publisher}{ACM}, \bibinfo{address}{New York, NY, USA},
  \bibinfo{pages}{2333--2338}.
\newblock
\showISBNx{978-1-4503-2263-8}


\bibitem[\protect\citeauthoryear{Hui, Yates, Berberich, and de~Melo}{Hui
  et~al\mbox{.}}{2017}]%
        {Hui-etal:2017:EMNLP}
\bibfield{author}{\bibinfo{person}{Kai Hui}, \bibinfo{person}{Andrew Yates},
  \bibinfo{person}{Klaus Berberich}, {and} \bibinfo{person}{Gerard de Melo}.}
  \bibinfo{year}{2017}\natexlab{}.
\newblock \showarticletitle{PACRR: A Position-Aware Neural IR Model for
  Relevance Matching}. In \bibinfo{booktitle}{\emph{Proceedings of the 2017
  Conference on Empirical Methods in Natural Language Processing}}.
  \bibinfo{publisher}{Association for Computational Linguistics},
  \bibinfo{pages}{1060--1069}.
\newblock
\urldef\tempurl%
\url{http://aclweb.org/anthology/D17-1111}
\showURL{%
\tempurl}


\bibitem[\protect\citeauthoryear{J\"{a}rvelin and
  Kek\"{a}l\"{a}inen}{J\"{a}rvelin and Kek\"{a}l\"{a}inen}{2000}]%
        {Jarvelin:2000}
\bibfield{author}{\bibinfo{person}{Kalervo J\"{a}rvelin} {and}
  \bibinfo{person}{Jaana Kek\"{a}l\"{a}inen}.} \bibinfo{year}{2000}\natexlab{}.
\newblock \showarticletitle{IR Evaluation Methods for Retrieving Highly
  Relevant Documents}. In \bibinfo{booktitle}{\emph{Proceedings of the 23rd
  Annual International ACM SIGIR Conference on Research and Development in
  Information Retrieval}} \emph{(\bibinfo{series}{SIGIR '00})}.
  \bibinfo{publisher}{ACM}, \bibinfo{address}{New York, NY, USA},
  \bibinfo{pages}{41--48}.
\newblock
\showISBNx{1-58113-226-3}


\bibitem[\protect\citeauthoryear{Joachims}{Joachims}{2002}]%
        {Joachims:2002b}
\bibfield{author}{\bibinfo{person}{Thorsten Joachims}.}
  \bibinfo{year}{2002}\natexlab{}.
\newblock \showarticletitle{Optimizing Search Engines Using Clickthrough Data}.
  In \bibinfo{booktitle}{\emph{Proceedings of the Eighth ACM SIGKDD
  International Conference on Knowledge Discovery and Data Mining}}
  \emph{(\bibinfo{series}{KDD '02})}. \bibinfo{publisher}{ACM},
  \bibinfo{address}{New York, NY, USA}, \bibinfo{pages}{133--142}.
\newblock
\showISBNx{1-58113-567-X}


\bibitem[\protect\citeauthoryear{Kim, Lu, and Wilbur}{Kim
  et~al\mbox{.}}{2015}]%
        {Kim-etal:2015}
\bibfield{author}{\bibinfo{person}{Sun Kim}, \bibinfo{person}{Zhiyong Lu},
  {and} \bibinfo{person}{W.~John Wilbur}.} \bibinfo{year}{2015}\natexlab{}.
\newblock \showarticletitle{Identifying named entities from PubMed® for
  enriching semantic categories}.
\newblock \bibinfo{journal}{\emph{BMC Bioinformatics}} \bibinfo{volume}{16},
  \bibinfo{number}{1} (\bibinfo{date}{21 Feb} \bibinfo{year}{2015}),
  \bibinfo{pages}{57}.
\newblock
\showISSN{1471-2105}
\urldef\tempurl%
\url{https://doi.org/10.1186/s12859-015-0487-2}
\showDOI{\tempurl}


\bibitem[\protect\citeauthoryear{Kusner, Sun, Kolkin, and Weinberger}{Kusner
  et~al\mbox{.}}{2015}]%
        {Kusner:2015:ICML}
\bibfield{author}{\bibinfo{person}{Matt Kusner}, \bibinfo{person}{Yu Sun},
  \bibinfo{person}{Nicholas Kolkin}, {and} \bibinfo{person}{Kilian
  Weinberger}.} \bibinfo{year}{2015}\natexlab{}.
\newblock \showarticletitle{From Word Embeddings To Document Distances}. In
  \bibinfo{booktitle}{\emph{Proceedings of The 32nd International Conference on
  Machine Learning}} \emph{(\bibinfo{series}{ICML 2015})}.
  \bibinfo{publisher}{JMLR}, \bibinfo{pages}{957–--966}.
\newblock


\bibitem[\protect\citeauthoryear{Lafferty and Zhai}{Lafferty and Zhai}{2001}]%
        {Lafferty:Zhai:2001}
\bibfield{author}{\bibinfo{person}{John Lafferty} {and}
  \bibinfo{person}{Chengxiang Zhai}.} \bibinfo{year}{2001}\natexlab{}.
\newblock \showarticletitle{Document Language Models, Query Models, and Risk
  Minimization for Information Retrieval}. In
  \bibinfo{booktitle}{\emph{Proceedings of the 24th Annual International ACM
  SIGIR Conference on Research and Development in Information Retrieval}}
  \emph{(\bibinfo{series}{SIGIR '01})}. \bibinfo{publisher}{ACM},
  \bibinfo{address}{New York, NY, USA}, \bibinfo{pages}{111--119}.
\newblock
\showISBNx{1-58113-331-6}
\urldef\tempurl%
\url{https://doi.org/10.1145/383952.383970}
\showDOI{\tempurl}


\bibitem[\protect\citeauthoryear{LeCun}{LeCun}{1989}]%
        {LeCun:1989}
\bibfield{author}{\bibinfo{person}{Yann LeCun}.}
  \bibinfo{year}{1989}\natexlab{}.
\newblock \showarticletitle{Generalization and Network Design Strategies}.
\newblock In \bibinfo{booktitle}{\emph{Connectionism in Perspective}},
  \bibfield{editor}{\bibinfo{person}{R.~Pfeifer},
  \bibinfo{person}{Z.~Schreter}, \bibinfo{person}{F.~Fogelman}, {and}
  \bibinfo{person}{L.~Steels}} (Eds.). \bibinfo{publisher}{Elsevier},
  \bibinfo{address}{Zurich, Switzerland}.
\newblock


\bibitem[\protect\citeauthoryear{Lu, Kim, and Wilbur}{Lu et~al\mbox{.}}{2009}]%
        {Lu:2009}
\bibfield{author}{\bibinfo{person}{Zhiyong Lu}, \bibinfo{person}{Won Kim},
  {and} \bibinfo{person}{W.~John Wilbur}.} \bibinfo{year}{2009}\natexlab{}.
\newblock \showarticletitle{Evaluation of query expansion using {MeSH} in
  {PubMed}}.
\newblock \bibinfo{journal}{\emph{Information retrieval}} \bibinfo{volume}{12},
  \bibinfo{number}{1} (\bibinfo{date}{May} \bibinfo{year}{2009}),
  \bibinfo{pages}{69--80}.
\newblock


\bibitem[\protect\citeauthoryear{Lu and Li}{Lu and Li}{2013}]%
        {Lu:Li:2013:NIPS}
\bibfield{author}{\bibinfo{person}{Zhengdong Lu} {and} \bibinfo{person}{Hang
  Li}.} \bibinfo{year}{2013}\natexlab{}.
\newblock \showarticletitle{A Deep Architecture for Matching Short Texts}.
\newblock In \bibinfo{booktitle}{\emph{Advances in NIPS 26}},
  \bibfield{editor}{\bibinfo{person}{C.~Burges}, \bibinfo{person}{L.~Bottou},
  \bibinfo{person}{M.~Welling}, \bibinfo{person}{Z.~Ghahramani}, {and}
  \bibinfo{person}{K.~Q. Weinberger}} (Eds.). \bibinfo{pages}{1367--1375}.
\newblock


\bibitem[\protect\citeauthoryear{Maas, Hannun, and Ng}{Maas
  et~al\mbox{.}}{2013}]%
        {Maas-etal:2013}
\bibfield{author}{\bibinfo{person}{Andrew~L. Maas}, \bibinfo{person}{Awni~Y.
  Hannun}, {and} \bibinfo{person}{Andrew~Y. Ng}.}
  \bibinfo{year}{2013}\natexlab{}.
\newblock \showarticletitle{Rectifier Nonlinearities Improve Neural Network
  Acoustic Models}. In \bibinfo{booktitle}{\emph{Proceedings of the ICML
  Workshop on Deep Learning for Audio, Speech, and Language Processing}}
  \emph{(\bibinfo{series}{WDLASL 2013})}.
\newblock


\bibitem[\protect\citeauthoryear{Mikolov, Chen, Corrado, and Dean}{Mikolov
  et~al\mbox{.}}{2013a}]%
        {Mikolov:2013:ICLR}
\bibfield{author}{\bibinfo{person}{Tomas Mikolov}, \bibinfo{person}{Kai Chen},
  \bibinfo{person}{Greg~S. Corrado}, {and} \bibinfo{person}{Jeffrey Dean}.}
  \bibinfo{year}{2013}\natexlab{a}.
\newblock \showarticletitle{Efficient Estimation of Word Representations in
  Vector Space}. In \bibinfo{booktitle}{\emph{Proceedings of the Workshop at
  ICLR 2013}}.
\newblock


\bibitem[\protect\citeauthoryear{Mikolov, Sutskever, Chen, Corrado, and
  Dean}{Mikolov et~al\mbox{.}}{2013b}]%
        {Mikolov:2013:NIPS}
\bibfield{author}{\bibinfo{person}{Tomas Mikolov}, \bibinfo{person}{Ilya
  Sutskever}, \bibinfo{person}{Kai Chen}, \bibinfo{person}{Greg~S. Corrado},
  {and} \bibinfo{person}{Jeff Dean}.} \bibinfo{year}{2013}\natexlab{b}.
\newblock \showarticletitle{Distributed Representations of Words and Phrases
  and their Compositionality}.
\newblock In \bibinfo{booktitle}{\emph{Advances in NIPS 26}},
  \bibfield{editor}{\bibinfo{person}{C.~J.~C. Burges},
  \bibinfo{person}{L.~Bottou}, \bibinfo{person}{M.~Welling},
  \bibinfo{person}{Z.~Ghahramani}, {and} \bibinfo{person}{K.~Q. Weinberger}}
  (Eds.). \bibinfo{pages}{3111--3119}.
\newblock


\bibitem[\protect\citeauthoryear{Miller, Leek, and Schwartz}{Miller
  et~al\mbox{.}}{1999}]%
        {Miller:1999}
\bibfield{author}{\bibinfo{person}{David R.~H. Miller}, \bibinfo{person}{Tim
  Leek}, {and} \bibinfo{person}{Richard~M. Schwartz}.}
  \bibinfo{year}{1999}\natexlab{}.
\newblock \showarticletitle{A Hidden Markov Model Information Retrieval
  System}. In \bibinfo{booktitle}{\emph{Proceedings of SIGIR 1999}}.
  \bibinfo{publisher}{ACM}, \bibinfo{address}{New York, NY, USA},
  \bibinfo{pages}{214--221}.
\newblock
\showISBNx{1-58113-096-1}


\bibitem[\protect\citeauthoryear{Mitra and Craswell}{Mitra and
  Craswell}{2017}]%
        {Mitra-Craswell:2017:NeuralIR}
\bibfield{author}{\bibinfo{person}{Bhaskar Mitra} {and} \bibinfo{person}{Nick
  Craswell}.} \bibinfo{year}{2017}\natexlab{}.
\newblock \showarticletitle{Neural Models for Information Retrieval}.
\newblock \bibinfo{journal}{\emph{CoRR}}  \bibinfo{volume}{abs/1705.01509}
  (\bibinfo{year}{2017}).
\newblock
\urldef\tempurl%
\url{https://arxiv.org/abs/1705.01509}
\showURL{%
\tempurl}


\bibitem[\protect\citeauthoryear{Mitra, Diaz, and Craswell}{Mitra
  et~al\mbox{.}}{2017}]%
        {Mitra-etal:2017:WWW}
\bibfield{author}{\bibinfo{person}{Bhaskar Mitra}, \bibinfo{person}{Fernando
  Diaz}, {and} \bibinfo{person}{Nick Craswell}.}
  \bibinfo{year}{2017}\natexlab{}.
\newblock \showarticletitle{Learning to Match Using Local and Distributed
  Representations of Text for Web Search}. In
  \bibinfo{booktitle}{\emph{Proceedings of the 26th International Conference on
  World Wide Web}} \emph{(\bibinfo{series}{WWW '17})}.
  \bibinfo{publisher}{International World Wide Web Conferences Steering
  Committee}, \bibinfo{address}{Republic and Canton of Geneva, Switzerland},
  \bibinfo{pages}{1291--1299}.
\newblock
\showISBNx{978-1-4503-4913-0}
\urldef\tempurl%
\url{https://doi.org/10.1145/3038912.3052579}
\showDOI{\tempurl}


\bibitem[\protect\citeauthoryear{Mohan, Fiorini, Kim, and Lu}{Mohan
  et~al\mbox{.}}{2017}]%
        {Mohan-etal:2017:BioNLP17}
\bibfield{author}{\bibinfo{person}{Sunil Mohan}, \bibinfo{person}{Nicolas
  Fiorini}, \bibinfo{person}{Sun Kim}, {and} \bibinfo{person}{Zhiyong Lu}.}
  \bibinfo{year}{2017}\natexlab{}.
\newblock \showarticletitle{Deep Learning for Biomedical Information Retrieval:
  Learning Textual Relevance from Click Logs}. In
  \bibinfo{booktitle}{\emph{BioNLP 2017}}. \bibinfo{publisher}{Association for
  Computational Linguistics}, \bibinfo{address}{Vancouver, Canada,},
  \bibinfo{pages}{222--231}.
\newblock
\urldef\tempurl%
\url{http://www.aclweb.org/anthology/W17-2328}
\showURL{%
\tempurl}


\bibitem[\protect\citeauthoryear{Nalisnick, Mitra, Craswell, and
  Caruana}{Nalisnick et~al\mbox{.}}{2016}]%
        {Nalisnick-etal:2016}
\bibfield{author}{\bibinfo{person}{Eric Nalisnick}, \bibinfo{person}{Bhaskar
  Mitra}, \bibinfo{person}{Nick Craswell}, {and} \bibinfo{person}{Rich
  Caruana}.} \bibinfo{year}{2016}\natexlab{}.
\newblock \showarticletitle{Improving Document Ranking with Dual Word
  Embeddings}. In \bibinfo{booktitle}{\emph{Proceedings of the 25th
  International Conference Companion on World Wide Web}}
  \emph{(\bibinfo{series}{WWW '16 Companion})}.
  \bibinfo{publisher}{International World Wide Web Conferences Steering
  Committee}, \bibinfo{address}{Republic and Canton of Geneva, Switzerland},
  \bibinfo{pages}{83--84}.
\newblock
\showISBNx{978-1-4503-4144-8}
\urldef\tempurl%
\url{https://doi.org/10.1145/2872518.2889361}
\showDOI{\tempurl}


\bibitem[\protect\citeauthoryear{Pang, Lan, Guo, Xu, Wan, and Cheng}{Pang
  et~al\mbox{.}}{2016}]%
        {Pang-etal:2016}
\bibfield{author}{\bibinfo{person}{Liang Pang}, \bibinfo{person}{Yanyan Lan},
  \bibinfo{person}{Jiafeng Guo}, \bibinfo{person}{Jun Xu},
  \bibinfo{person}{Shengxian Wan}, {and} \bibinfo{person}{Xueqi Cheng}.}
  \bibinfo{year}{2016}\natexlab{}.
\newblock \bibinfo{title}{Text Matching as Image Recognition}.
\newblock   (\bibinfo{year}{2016}).
\newblock
\urldef\tempurl%
\url{https://www.aaai.org/ocs/index.php/AAAI/AAAI16/paper/view/11895}
\showURL{%
\tempurl}


\bibitem[\protect\citeauthoryear{Robertson, Walker, Jones, Hancock-Beaulieu,
  and Gatford}{Robertson et~al\mbox{.}}{1994}]%
        {Robertson:1994}
\bibfield{author}{\bibinfo{person}{Stephen~E. Robertson},
  \bibinfo{person}{Steve Walker}, \bibinfo{person}{Susan Jones},
  \bibinfo{person}{Micheline Hancock-Beaulieu}, {and} \bibinfo{person}{Mike
  Gatford}.} \bibinfo{year}{1994}\natexlab{}.
\newblock \showarticletitle{Okapi at {TREC-3}}. In
  \bibinfo{booktitle}{\emph{Proceedings of TREC 1994}}.
  \bibinfo{publisher}{NIST, Dept. of Commerce}, \bibinfo{pages}{109--126}.
\newblock


\bibitem[\protect\citeauthoryear{Rubner, Tomasi, and Guibas}{Rubner
  et~al\mbox{.}}{1998}]%
        {Rubner:1998}
\bibfield{author}{\bibinfo{person}{Y. Rubner}, \bibinfo{person}{C. Tomasi},
  {and} \bibinfo{person}{L.~J. Guibas}.} \bibinfo{year}{1998}\natexlab{}.
\newblock \showarticletitle{A metric for distributions with applications to
  image databases}. In \bibinfo{booktitle}{\emph{Proceedings of the Sixth
  International Conference on Computer Vision}}. \bibinfo{publisher}{IEEE}.
\newblock
\showISBNx{81-7319-221-9}


\bibitem[\protect\citeauthoryear{Severyn and Moschitti}{Severyn and
  Moschitti}{2015}]%
        {Severyn:2015}
\bibfield{author}{\bibinfo{person}{Aliaksei Severyn} {and}
  \bibinfo{person}{Alessandro Moschitti}.} \bibinfo{year}{2015}\natexlab{}.
\newblock \showarticletitle{Learning to Rank Short Text Pairs with
  Convolutional Deep Neural Networks}. In \bibinfo{booktitle}{\emph{Proceedings
  of SIGIR 2015}}. \bibinfo{publisher}{ACM}, \bibinfo{address}{New York, NY,
  USA}, \bibinfo{pages}{373--382}.
\newblock
\showISBNx{978-1-4503-3621-5}


\bibitem[\protect\citeauthoryear{Shen, He, Gao, Deng, and Mesnil}{Shen
  et~al\mbox{.}}{2014}]%
        {Shen:2014}
\bibfield{author}{\bibinfo{person}{Yelong Shen}, \bibinfo{person}{Xiaodong He},
  \bibinfo{person}{Jianfeng Gao}, \bibinfo{person}{Li Deng}, {and}
  \bibinfo{person}{Gr{\'e}goire Mesnil}.} \bibinfo{year}{2014}\natexlab{}.
\newblock \showarticletitle{A Latent Semantic Model with Convolutional-Pooling
  Structure for Information Retrieval}. In
  \bibinfo{booktitle}{\emph{Proceedings of CIKM 2014}}.
  \bibinfo{publisher}{ACM}, \bibinfo{address}{New York, NY, USA},
  \bibinfo{pages}{101--110}.
\newblock
\showISBNx{978-1-4503-2598-1}


\bibitem[\protect\citeauthoryear{Urban, Geras, Kahou, Aslan, Wang, Mohamed,
  Philipose, Richardson, and Caruana}{Urban et~al\mbox{.}}{2017}]%
        {Urban-etal:2017}
\bibfield{author}{\bibinfo{person}{Gregor Urban}, \bibinfo{person}{Krzysztof~J.
  Geras}, \bibinfo{person}{Samira~Ebrahimi Kahou}, \bibinfo{person}{Ozlem
  Aslan}, \bibinfo{person}{Shengjie Wang}, \bibinfo{person}{Abdelrahman
  Mohamed}, \bibinfo{person}{Matthai Philipose}, \bibinfo{person}{Matt
  Richardson}, {and} \bibinfo{person}{Rich Caruana}.}
  \bibinfo{year}{2017}\natexlab{}.
\newblock \showarticletitle{Do Deep Convolutional Nets Really Need to be Deep
  and Convolutional?}. In \bibinfo{booktitle}{\emph{Proceedings of the Fifth
  International Conference on Learning Representations}}
  \emph{(\bibinfo{series}{ICLR 2017})}. \bibinfo{address}{Toulon, France}.
\newblock


\bibitem[\protect\citeauthoryear{Xiong, Dai, Callan, Liu, and Power}{Xiong
  et~al\mbox{.}}{2017}]%
        {Xiong-etal:2017}
\bibfield{author}{\bibinfo{person}{Chenyan Xiong}, \bibinfo{person}{Zhuyun
  Dai}, \bibinfo{person}{Jamie Callan}, \bibinfo{person}{Zhiyuan Liu}, {and}
  \bibinfo{person}{Russell Power}.} \bibinfo{year}{2017}\natexlab{}.
\newblock \showarticletitle{End-to-End Neural Ad-hoc Ranking with Kernel
  Pooling}. In \bibinfo{booktitle}{\emph{Proceedings of the 40th International
  ACM SIGIR Conference on Research and Development in Information Retrieval}}
  \emph{(\bibinfo{series}{SIGIR '17})}. \bibinfo{publisher}{ACM},
  \bibinfo{address}{New York, NY, USA}, \bibinfo{pages}{55--64}.
\newblock
\showISBNx{978-1-4503-5022-8}
\urldef\tempurl%
\url{https://doi.org/10.1145/3077136.3080809}
\showDOI{\tempurl}


\bibitem[\protect\citeauthoryear{Zhai and Lafferty}{Zhai and Lafferty}{2004}]%
        {Zhai:Lafferty:2004}
\bibfield{author}{\bibinfo{person}{Chengxiang Zhai} {and} \bibinfo{person}{John
  Lafferty}.} \bibinfo{year}{2004}\natexlab{}.
\newblock \showarticletitle{A Study of Smoothing Methods for Language Models
  Applied to Information Retrieval}.
\newblock \bibinfo{journal}{\emph{ACM Trans. Inf. Syst.}} \bibinfo{volume}{22},
  \bibinfo{number}{2} (\bibinfo{date}{April} \bibinfo{year}{2004}),
  \bibinfo{pages}{179--214}.
\newblock
\showISSN{1046-8188}


\bibitem[\protect\citeauthoryear{Zhang, Rahman, Braylan, Dang, Chang, Kim,
  McNamara, Angert, Banner, Khetan, McDonnell, Nguyen, Xu, Wallace, and
  Lease}{Zhang et~al\mbox{.}}{2016}]%
        {Zhang-etal:2016:NeuralIR-Review}
\bibfield{author}{\bibinfo{person}{Y. Zhang}, \bibinfo{person}{M.~M. Rahman},
  \bibinfo{person}{A. Braylan}, \bibinfo{person}{B. Dang}, \bibinfo{person}{H.
  Chang}, \bibinfo{person}{H. Kim}, \bibinfo{person}{Q. McNamara},
  \bibinfo{person}{A. Angert}, \bibinfo{person}{E. Banner}, \bibinfo{person}{V.
  Khetan}, \bibinfo{person}{T. McDonnell}, \bibinfo{person}{A.~T. Nguyen},
  \bibinfo{person}{D. Xu}, \bibinfo{person}{B.~C. Wallace}, {and}
  \bibinfo{person}{M. Lease}.} \bibinfo{year}{2016}\natexlab{}.
\newblock \showarticletitle{Neural Information Retrieval: {A} Literature
  Review}.
\newblock \bibinfo{journal}{\emph{CoRR}}  \bibinfo{volume}{abs/1611.06792}
  (\bibinfo{year}{2016}).
\newblock
\urldef\tempurl%
\url{http://arxiv.org/abs/1611.06792}
\showURL{%
\tempurl}


\end{thebibliography}
